\documentclass[%
 reprint,
superscriptaddress,
 amsmath,amssymb,
 aps,
pre,
]{revtex4-2}

\usepackage{appendix}
\usepackage{epsfig,dsfont,amssymb,amsmath,amsthm,amsfonts,amsbsy,mathrsfs}
\usepackage{graphicx}
\usepackage{color}
\usepackage{bm}
\usepackage{multirow}
\usepackage{natbib}
\usepackage{tikz}
\usetikzlibrary{arrows,shapes,chains}

\newcommand{\blue}[1]{{\textcolor{black}{#1}}}

\newcommand{\rv}[1]{{\bf r}}
\newcommand{\qv}[1]{{\bf q}}

\newcommand{\uv}[1]{{ \underline{ #1}}}

\newcommand{\beq}{\begin{equation}}
\newcommand{\eeq}{\end{equation}}
\newcommand{\bpm}{\begin{pmatrix}}
\newcommand{\epm}{\end{pmatrix}}
\newcommand{\p}{\partial}

\begin{document}
\preprint{APS/123-QED}

\title{Non-equilibrium phase transitions in hybrid Voronoi models of cell colonies}

\author{Mattia Miotto}
\affiliation{Center for Life Nano \& Neuro Science, Italian Institute of Technology, Viale Regina Elena 261, Rome, Italy.}

\author{Giancarlo Ruocco}
\affiliation{Center for Life Nano \& Neuro Science, Italian Institute of Technology, Viale Regina Elena 261, Rome, Italy.}
\affiliation{Department of Physics, Sapienza University of Rome, Piazzale Aldo Moro, Rome, Italy.}

\author{Matteo Paoluzzi}
\affiliation{Department of Physics, Sapienza University of Rome, Piazzale Aldo Moro, Rome, Italy.}

\begin{abstract}
Eukaryotic cells are characterized by a stiff nucleus whose role in governing the collective behavior of cell aggregates is often underestimated. However, increasing experimental evidence links nuclear mechanics to phenotypic transitions, such as the epithelial-to-mesenchymal transition (EMT). In this work, we explore the effect of short-range repulsive forces on the non-equilibrium dynamics of the self-propelled Voronoi model. We demonstrate that the competition between steric repulsion (representing nuclear/cellular compressibility) and vertex interactions (mimicking cell-cell adhesion and cytoskeleton organization) generates a variety of non-equilibrium phase transitions, ranging from Motility-Induced Phase Separation (MIPS) and mesenchymal-like phases to disordered \blue{dense} configurations. Notably, we find that tuning the effective size or compressibility of the nucleus provides an additional pathway to cross phase boundaries, consistent with experimental observations.
\end{abstract}

\maketitle

\section{Introduction}
Biological tissues are \blue{compact} collections of cells \blue{at high densities} \cite{alert2020physical,trepat2018mesoscale,Bi2016,henkes2020dense}. Each cell is a deformable object whose shape can fluctuate \blue{and change due to interactions} with \blue{other cells}. Various models incorporate some of those minimal ingredients for capturing the collective behavior of cells ~\cite{alert2020physical,trepat2018mesoscale,camley2017physical,barton2017active, Li2018} in different situations ranging from morphogenesis to cancer invasion ~\cite{Miotto2023,sunyer2016collective, Friedl17, Miotto2021}.
Depending on the biological situation of interest, different coarse-graining descriptions of cells might be very effective. These descriptions range from particle-based models \cite{Smeets16,henkes2020dense,giavazzi2017giant,Paoluzzi2022, paoluzzi2024flocking}, phase field models \cite{PhysRevLett.105.108104,PhysRevE.104.054410,chiang2024intercellular,PhysRevE.98.042402, PhysRevLett.129.148101,PhysRevLett.125.038003,PhysRevLett.122.048004}, Potts models \cite{PhysRevLett.69.2013,nandi21,PhysRevLett.132.248401,PhysRevE.47.2128}, and Voronoi/Vertex models \cite{nagai2001dynamic,bi2015density,Bi2016,giavazzi2018flocking,merkel2018geometrically,merkel2019minimal,erdemci2021effect, PhysRevResearch.2.043026, yang2017correlating,farhadifar2007influence}.
\blue{Approaches based on ring polymers provide another paradigm for modeling single cell \cite{paoluzzi2016shape}, but also cell collectives \cite{gnan2019microscopic,PhysRevResearch.6.L012036}.}
A single cell is characterized by \blue{hierarchically arranged structures.}
For instance, cell shape fluctuations can be included as a leading ingredient necessary for modeling \blue{cell aggregates} more realistically.
Such fluctuations have been found to act as a driving force for jamming/unjamming transition~\cite{Park15,Arora2024,bi2015density,Bi2016,Sadhukhan2024}. 
Experimental studies revealed the importance of non-equilibrium glass-like or jamming-like transition in confluent monolayers \cite{Park15,Malinverno17,kang2021novel,Arora2024}.

\blue{While cells can adapt to different environments and move across barriers by changing their shape~\cite{Saito2024}, the presence of a stiff nucleus sets a fundamental limit to their deformability. Despite its importance, the role of nuclear stiffness
is typically neglected in standard Voronoi and Vertex models, where cells are treated as fully deformable tiles. 
In this framework, introducing a short-range nuclear repulsion into the Voronoi formalisms is crucial to understanding how nuclear hardness impacts collective cell behavior (see, for instance, the reviews~\cite{CaleroCuenca2018, Boutillon2024}). Experimental evidence supports this need: measuring migration speeds and cell/nucleus shapes, Wolf \textit{et al.}~\cite{Wolf2013} identified the deformability of the nucleus as one of the first-order rate-limiting physicochemical determinants of cell migration.}

Grosser \textit{et al.}~\cite{Grosser2021} compared spheroids of cancerous and noncancerous cell lines via bulk experiments and/or single live-cell tracking, finding that cancerous cells populations are fluidized by active cells moving through the \blue{epithelial} tissue, with the degree of tissue fluidity correlating with elongated cell and nucleus shapes. They speculate that individual cell and nucleus shape may serve as a marker for metastatic potential.
\blue{The effect of cell nucleus in single cell simulations within the phase field model frameworks have been studied in ~\cite{Chojowski2024}.}
In other words, 
\blue{However,} whether and how the presence of a stiff nucleus impacts the collective behavior of \blue{cells}
remains an open question.

Here, we introduce a hybrid \blue{version of the} Voronoi model \blue{of biological tissue} that \blue{incorporates the presence of a stiff nucleus, modeled through a power-law soft repulsive potential.}
Our hybrid Voronoi model allows us to describe cells in both mesenchymal \blue{phases, characterized by negligible geometrical contributions to the energy}
, and epithelial phases~ \cite{Mitchel2020} \blue{where the vertex-based interactions dominate over short-range repulsion.} Indeed, by varying 
the relative strength between \blue{the} Vertex configurational energy and short-range steric interactions,
we interpolate from a pure particle-based model to \blue{pure Voronoi model.}
Note that our approach incorporates only a few parameters for controlling different ingredients, \blue{such} as typical shape elongation, cell motility, and nuclear stiffness.
In contrast, other approaches require
the fine-tuning of many parameters and is quite expensive from the numerical point of view \cite{alert2020physical}.
In the following, we show that the system undergoes several non-equilibrium phase transitions from the so-called Motility-Induced Phase Separation (MIPS) \cite{Tailleur08}, to standard active liquid (\blue{when the repulsive interactions dominate over the geometrical ones}), to \blue{what we call}
a Voronoi \blue{fluid, i.e., when the short-range repulsion is a perturbation with respect to the geometrical interaction}, and a disordered solid.
\blue{As control parameters that drive the system to the different phases, we tune the relative weight of the repulsive forces over the geometrical Vertex energy, typical shape elongation, cell motility, and nuclear hardness.}

\begin{figure*}[t]
\centering
\includegraphics[width=.95\textwidth]{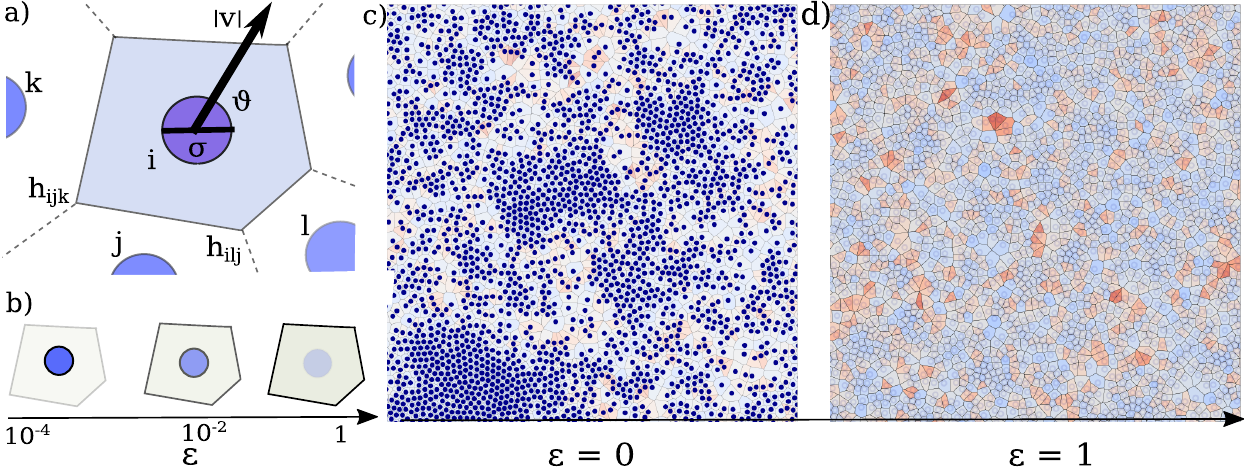}
\caption{
\textbf{Model of cell colony.}
\textbf{a)} Schematic representation of a cell. The effective surface of the cells is given by a Voronoi tessellation of the 2D plane, \blue{which} is defined by the vertices, h, shared by triplets of neighboring cells.
\blue{The} cell nucleus is rendered via the repulsive interaction and has an effective diameter of $\sigma$. The vector represents the self-propulsion direction of modulus $|\dot{\boldsymbol{r}^i}|$. 
 \textbf{b)} Schematic representation of the trade-off between the repulsive potential energy and the Voronoi interaction acting among cell nuclei. The relative intensity of the two terms is modulated by the $\epsilon$ parameter introduced in Eq.~\ref{eq:tot_force}. \textbf{c)} Snapshot in the pure repulsive regime with a long persistence time. Cell nuclei are represented as blue circles. The corresponding Voronoi tesselation is shown on the background with shaded colors.
 \textbf{d)} Snapshot in the pure Voronoi regime \blue{$\epsilon=1$}. 
Cell colors range from dark blue to dark red as the cell shape factor increases.}
\label{fig:1}
\end{figure*}

\begin{figure*}[t]
\centering
\includegraphics[width=.95\textwidth]{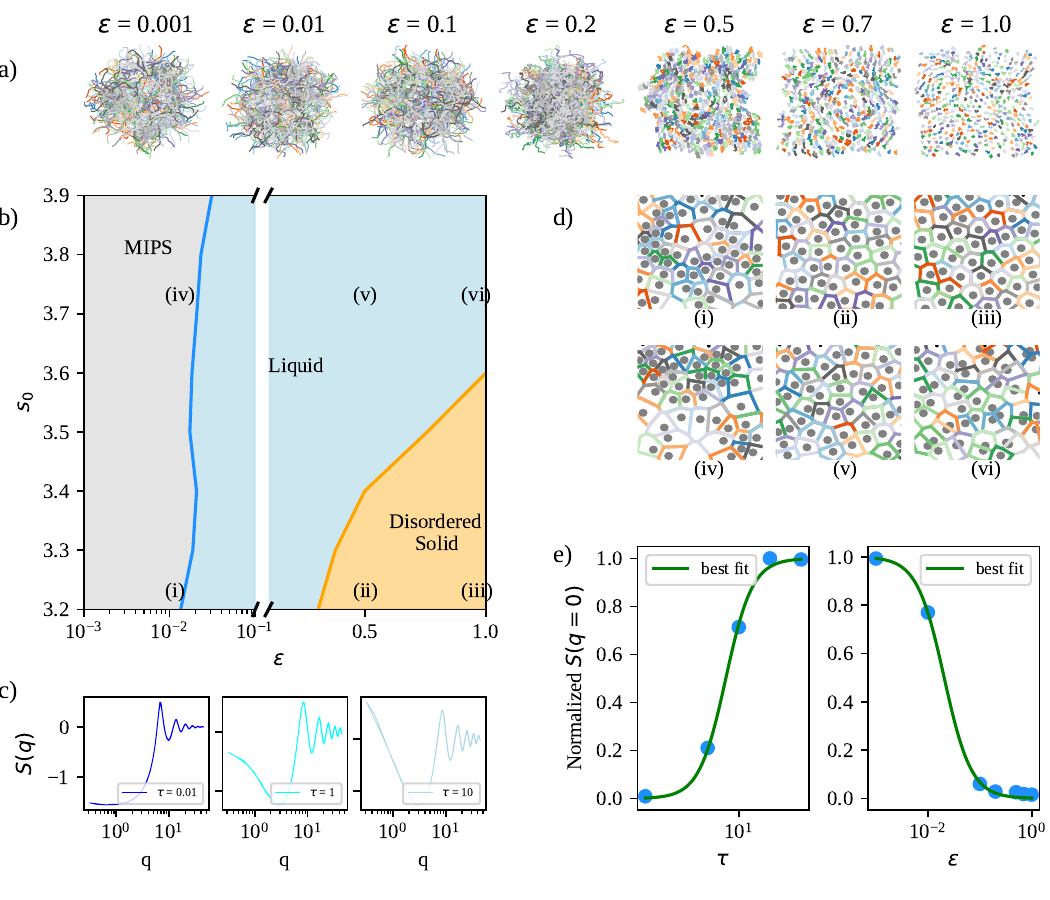}
\caption{\textbf{Competition between nuclear short-range repulsion and geometrical forces.}
\textbf{(a)}  \blue{Representative trajectories of 400 cells upon increasing the strength of the geometrical force, while keeping fixed the persistence time, target perimeter and self propulsion velocity to $\tau = 1000$, $p_0 = 3.2$, and $v_0 = 0.5$, respectively.} Each trajectory is marked with a different color. \textbf{(b)} Phase diagram as a function of $s_0$ and $\epsilon$,\blue{keeping fixed $\tau = 1000$ and $v_0 = 0.5$}.
\textbf{(c)} Static structure factor $S(q)$ for different values of the persistence time, $\tau$ in the limit of pure nuclear repulsion ($\epsilon\!=\!0$). \textbf{(d)} Voronoi tessellation of a set of representative zoomed snapshots of the population in the stationary regime for different choices of $\epsilon$ and $s_0$, marked in the phase diagram reported in panel b). \textbf{(e)} Normalized $S(q)$ at long distances ($q = 0$)  as a function of the persistence time in \blue{the pure nuclear repulsion regime  for $\epsilon = 0.001$ (left) and} as a function of the relative strength of the geometrical force for a choice of the persistence time, $\tau=10^3$ (right). Blue dots mark the value extracted from simulations, while the green line corresponds to the best-fit solution of the sigmoidal function, $y/y_{max}= 1/(1+e^{-x/x_0})$, reported in Figure~\ref{sfig:s1}. Flex points mark the boundaries of the MIPS.} 
\label{fig:2}
\end{figure*}

\begin{figure*}[t]
\centering
\includegraphics[width=.95\textwidth]{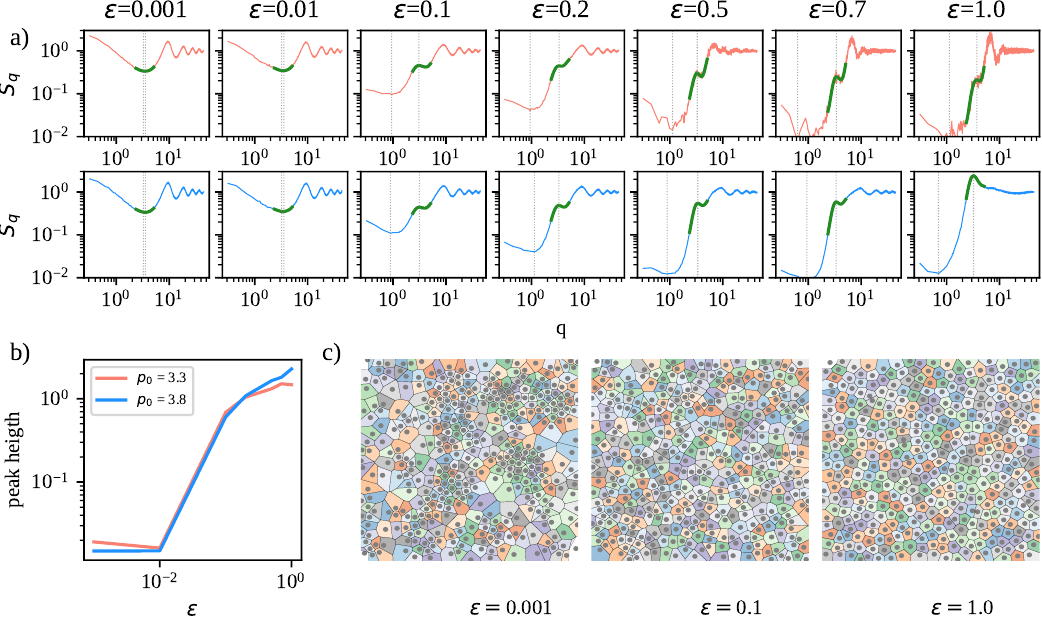}
\caption{\textbf{\blue{Structural properties.}} a) $S(q)$ as a function of the relative strength between repulsive and geometrical forces, $\epsilon$, for two choices of the shape factor $s_0$. The top row (respectively bottom) reports results for $s_0 = 3.3$ ($s_0 = 3.8$). Green line \blue{marks} the region corresponding to the expected Voronoi peak. \textbf{b)} Difference between the maximum and minimum values of $S(q)$ in the region of the Voronoi peak highlighted in panel a) as a function of $\epsilon$. \textbf{c)} Voronoi tesselation of snapshots taken from the stationary phase at three different values of the $\epsilon$ parameter. \blue{All other control parameters are kept fixed to $\tau = 1000$, $v_0 = 0.5$, and $\sigma_0 = 0.25$.}}
\label{fig:3}
\end{figure*}

\section{Hybrid Voronoi Model}
We consider 
a hybrid version of the self-propelled Voronoi model \cite{Bi2016,barton2017active}, that is built as follows.
The $N$ cells are confined into a square box of side $L\!=\!\sqrt{N}$ with periodic boundary conditions. We indicate with ${\boldsymbol{r}}^i$ the position of the $i-$th cell nucleus center (with $i\!=\!1,..,N$) whose equation of motion reads
\begin{equation}
    \label{eq:dotr}
    \dot{\boldsymbol{r}}^i = v_0 \hat{e}_i + \mu \boldsymbol{F}_i 
\end{equation}
where $\boldsymbol{F}_i$ is the total force acting on the cell, $\mu$ the mobility,  
and  $v_0$  the modulus of the self-propulsion velocity whose direction in 2D is given by  $\hat{e}_i$ that,in 2D, is parameterized by the angle $\theta_i$, i.e.,  $\hat{e}_i = (\cos\theta_i, \sin\theta_i)$. 
In writing (\ref{eq:dotr}), we consider the case where each cell moves with same self-propulsion velocity $v_0$. Cell mobility $\mu$ is set to one. 
In the following, cell nuclei are taken as the centers of polygons in a Voronoi tessellation of the space.
Cell centers undergo Active Brownian dynamics~\cite{Bechinger2016} in which the orientation of the self-propulsion
$\theta_i$ follows a standard Langevin equation of the form $\dot{\theta}_i = \eta_i$, with  $\langle \eta_i \rangle = 0$ and $\langle\eta_i(t)\eta_j(s) \rangle = \frac{2}{\tau}\delta_{ij}\delta(t-s)$. 
Here, we introduce the persistence time of the active motion, $\tau$ that is the inverse of the rotational diffusivity coefficient $D_r$ ~\cite{Bechinger2016,fodor2018statistical,marchetti2013hydrodynamics}.

The total mechanical force $\boldsymbol{F}_i$ acting on each cell can be decomposed into two contributions: one governing cell shape fluctuations and the other enforcing nuclear incompressibility
\begin{equation}
\label{eq:tot_force}
    \boldsymbol{F}_i = -\nabla_{\boldsymbol{r}_i} \left[ \epsilon E_{v} + (1-\epsilon) \Phi\right]
\end{equation}
where $\epsilon$ is a parameter tuning the relative intensities of geometrical $E_v$ and nuclear repulsion $\Phi$ energies. 
Note that for $\epsilon\!=\!0$ one recovers \blue{an active soft-disk model,} while for $\epsilon\!=\!1$ the system behaves as the standard Voronoi model. In Fig. (\ref{fig:1})a-b, we depict the relevant features of the model.
In particular, the term  $E_v \!=\! \frac{1}{2} \sum_i^N E_v^i$ is the geometrical energetic contribution given by the Vertex energy \cite{nagai2001dynamic,farhadifar2007influence,hufnagel2007mechanism,staple2010mechanics,hilgenfeldt2008physical,manning2010coaction,wang2012regional,chiou2012mechanical,fletcher2014vertex,bi2015density} $E_v^i \!=\!  K_p (p_i -p_0)^2 + K_a (A_i -A_0)^2 $
in which  $p_i$ and  $A_i$ represent the perimeter and area of cell $i$, respectively; $K_p$, $K_a$  are constants depending on the specific cell type.
The area modulus $K_a$ , 
determines how resistant a cell is to deviations from its preferred area. It penalizes deviations of the actual area of the cell from its target area $A_0$. The latter reflects the size that cells tend to maintain due to internal forces such as pressure, growth constraints, or packing density. Similarly,  $K_p$ is the perimeter elasticity constant or perimeter modulus, which controls how resistant a cell is to changes in its perimeter..
Analogous to 
$A_0$, $p_0$ reflects the cell’s preferred shape in terms of its perimeter and typically accounts for the contractility of the actomyosin network at the cell boundary.
In the following, we set $K_p\!=\!K_a\!=\!A_0\!=\!1$ changing the shape parameter $s_0 \!\equiv\! p_0 / \sqrt{A_0}$ \blue{for tuning the typical cell elongation \cite{bi2015density,Bi2016,giavazzi2018flocking}.}

\begin{figure*}[t]
\centering
\includegraphics[width=.95\textwidth]{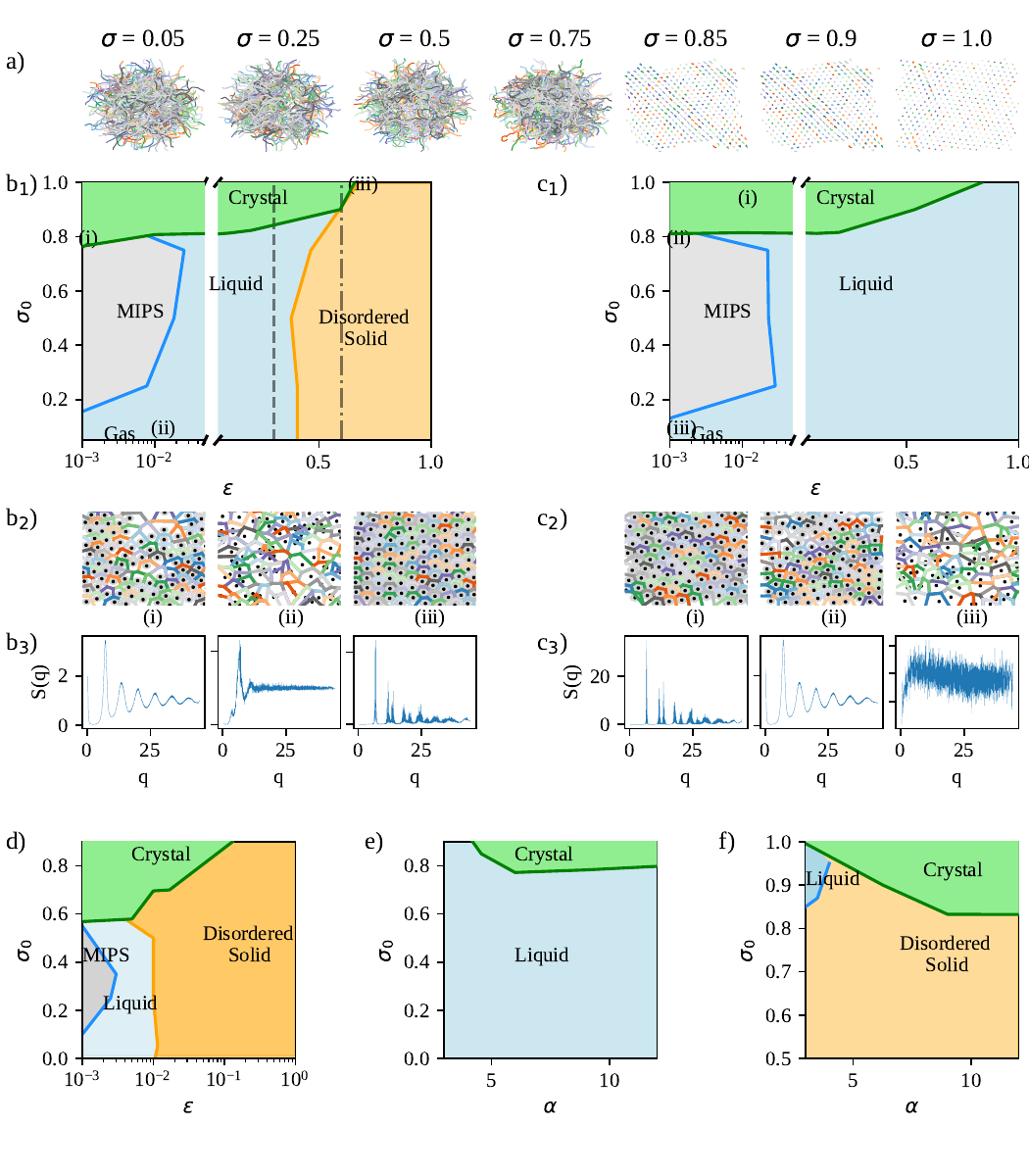}
\caption{\textbf{Nucleus size as a driver for the Epithelial-to-Mesenchymal transition.} \textbf{a)} Representative trajectories observed increasing the \blue{nuclear} size in the presence of weak geometrical forces.
\textbf{b${}_1$)} Phase diagram of the behavior of cell colonies as a function of $\epsilon$ and of nuclei size, $\sigma_0$.  Lines mark the borders between the different observed phases ($s_0=3.3$).  \blue{Dashed and dashed-dotted lines mark the $\epsilon$ values used to simulate results shown in panels e) and f), respectively.}
\textbf{b${}_2$)} Voronoi tesselation of three snapshots at different values of $\epsilon$ and $\sigma_0$, as shown in panel b1).
\textbf{b${}_3$)} $S(q)$ for the three sets of parameters of snapshots in panel b${}_2$).\textbf{c${}_{1-3}$} Same as in panels b${}_{1-3}$) but considering cells with $s_0=3.8$.
\textbf{d)} Phase diagram of the behavior of cell colonies as a function of $\epsilon$ and of \blue{nuclear} size, $\sigma_0$, for $s_0=3.3$ and $v_0 = 0.01$.
\textbf{e)} Phase diagram of the behavior of cell colonies as a function of $\alpha$ and of \blue{nuclear} size, $\sigma_0$, for $s_0=3.3$, $v_0 = 0.5$, and $\epsilon = 0.3$. \textbf{f)} Same as in e) but for $\epsilon = 0.6$. \blue{ Boundary lines in the plots correspond to the best-fit solution of the sigmoidal function, $y/y_{max}= 1/(1+e^{-x/x_0})$ as reported in Figure~\ref{sfig:s2} and~\ref{sfig:s3}.}}
\label{fig:4}
\end{figure*}

$\Phi$ accounts for an additional key feature of cell architecture, i.e. the steric hindrance exerted by cell nuclei, whose \blue{hardness} ultimately limits the compressibility of cell layers. To this aim, we inserted a  repulsive interaction between cell nuclei  given by $\Phi = \sum_{i < j} \phi(r_{ij})$ and $\phi(r) = \left( \frac{\sigma_0}{r}\right)^{\alpha}$,
where $\sigma_0$ gives an effective size for the cell nucleus, while $\alpha$ quantifies the nuclear \blue{hardness}/compressibility.
We emphasize that, despite a power law form of the interacting potential, it is suitable for describing excluded volume effects due to short-range stiff forces, indeed this is the typical interaction one considers in modeling colloidal particles in both, equilibrium and non-equilibrium models.
The interaction is suitable for modeling very short-range and strong repulsive forces because of the large value of the exponent combined with a short cutoff distance \blue{$r_{cut}=\sqrt{3} \sigma_0$}, 
in computing the force.
We thus combine repulsive interactions, given by the excluded volume effect due to \blue{the nuclear hardness} to the Vertex energy term, which promotes a spatial configuration of the system described by a Voronoi tessellation.  The vertex energy considers mechanical interactions due to cell shape fluctuations in a regime where each cell is surrounded by neighboring cells. This is a typical scenario for confluent \blue{epithelial} tissue. Note that any fluctuation (weakly) preserving cell area is allowed within Voronoi models. 
Our model allows us to explore the trade-off between the behavior of highly deformable cells, where excluded volume is dominated by the presence of the nucleus, and the regime in which the whole cell has a relevant stiffness.

\blue{
To document and quantify the transitions between different stationary states, we monitored several observables, as detailed in the Appendices. First, through the static structure factor $S(q)$, we (i) identify the presence of Motility-Induced Phase Separation (MIPS) and (ii) quantify the structural properties of the liquid phase. The onset of phase separation is characterized by an increase in $S(0) = \lim_{q \to 0} S(q)$. Due to the finite size of the simulation box, $S(0)$ is estimated by measuring the magnitude of $S(q_{min})$, where $q_{min} = 2\pi / L$ is the smallest wavevector allowed by the boundary conditions. Furthermore, changes in the local structure of the fluid phase are captured by the evolution of the first peak of $S(q)$. To investigate the emergence of a crystalline hexatic phase, we compute the hexatic order parameter $\Psi_6$. Finally, to document the liquid-to-solid transition, we measure the mean square displacement $MSD(t)$. The transition is quantified by the effective diffusion coefficient $D_{eff}$, obtained from the long-time limit $D_{eff} = \lim_{t \to \infty} MSD(t) / 4t$.
}

\section{Competition between excluded volume and geometrical forces}
\blue{MIPS represents a hallmark non-equilibrium transition in scalar active matter, driven by self-propelled motion that lacks any equilibrium counterpart. Since persistent random motion characterizes epithelial tissues \cite{Malinverno17} and MIPS has been observed even in dense phases \cite{Paoluzzi2022}, it is crucial to understand how additional interactions, specifically those of a purely geometric origin, as in Voronoi models, affect the stability of such active phase separation.
We thus start with exploring the impact of the mechanical forces due to $E_v$ on MIPS. To do so, we explore a phase diagram moving in $s_0$, i.e., the control parameter of the rigidity transition, and $\epsilon$, i.e., the relative weight between short-range repulsion and Vertex interactions.}

As a first result, as sketched in Figure~\ref{fig:1}c-d, we observe MIPS at a very small $\epsilon$ value. To explore systematically the interplay between \blue{the short-range repulsive interactions and the geometrical force due to the Vertex mechanical energy,}
we considered a homogeneous population of cells with a nucleus' effective radius of $\sigma_0\!=\!0.5$, self-propulsion velocity of $v_0\!=\!0.5$, \blue{persistence time of the active motion $\tau=10$}, $\alpha = 12$,  and shape parameter, $s_0\!=\!3.2$. 
We study the non-equilibrium steady state behavior of the system upon varying the relative strength of the two potential terms via tuning $\epsilon$. As one can see from the trajectories of the cell centers displayed in Figure~\ref{fig:2}a, the pure Voronoi regime \blue{$\epsilon=1$} is associated with a solid-like motion of the cells, whose centers oscillate around quasi-fixed positions, in agreement with early studies on the self-propelled Voronoi model \cite{Bi2016}.
Increasing the contribution of the nuclear repulsion, the system behaves more like a fluid, with trajectories having higher displacements. In particular, the effective diffusion coefficient~\cite{Bi2016}, $D_{eff}$ presents a sigmoidal trend as a function of $\epsilon$ associated with a solid-to-fluid phase transition of the system (see Appendix). Moreover, the critical $\epsilon$ increases as a function of the cell shape factor (orange curve in Fig.~\ref{fig:2}b) and reaches the limiting value of $1$ for $s_0\!\sim\!3.6$, as one would expect since the pure Voronoi model is known to exhibit a disordered solid to fluid transition upon varying the cell shape parameter~\cite{Bi2016}. 

Next, we analyzed the behavior of the cell population in the fluid region of the phase diagram. In the pure repulsive regime ($\epsilon\!=\!0 $), cells act like repulsive active particles of size $\sigma_0$. Such systems are known to exhibit MIPS as a function of $\tau$ \cite{fodor2018statistical}.  The transition between a homogeneous gas phase \blue{and} a mixture of a low-density fluid-like phase and an ordered crystal-like phase as a function of $\tau$ can be quantitatively measured by looking at the static structure factor $S(q)$,  at low $q$ vectors. In particular, Fig.~\ref{fig:2}c shows the behavior of $S(q)$ for three different values of $\tau$. As one can see, $S(q)$ at long distances increases with $\tau$ following a sigmoidal curve with critical $\tau\sim 10$ (see Fig.~\ref{fig:2}e).  
Looking at the low $q$ value of $S(q)$ as a function of the relative strength $\epsilon$ and $s_0$, we observe a phase transition between MIPS and standard active liquid at $\epsilon\!\sim\!0.02$ as shown in Fig.~\ref{fig:2}e and the blue line in Fig.~\ref{fig:2}b. 

\blue{The phase diagram documents two main findings: (i) The MIPS-fluid transition, i.e., the disappearance of active separation because of the geometrical interaction, is mostly independent of $s_0$, (ii) The transition from fluid to disordered solid, driven by the geometric interaction, produces a transition line $s_0(\epsilon)$, with $s_0(\epsilon)$ an increasing function of $\epsilon$, i.e., for $\epsilon < 1$ the Voronoi rigidity transition requires smaller $s_0$ values to occur. This finding indicates that short-range repulsive interactions disfavor the Voronoi rigidity transition, effectively expanding the fluid regime.}

Further analyzing the behavior of $S(q)$ \blue{to quantify the structural properties of the system}, we observe that besides the trend at $q = 0$, carrying information on MIPS, and the oscillating trend at high $q$, characteristic of fluids, the curves present a peculiar behavior at intermediate values of $q_V\sim 3$, \blue{Here, the static structure factor undergoes a qualitative change as $\epsilon$ increases.} 
\blue{We observe that, as MIPS disappears above a threshold value $\epsilon_{MIPS}$, the system changes its local structure especially over a length scale of order $1/q_V$. Specifically, $S(q)$ develops a secondary peak that is initially faint as $S(0)$ vanishes, remaining subdominant to the primary peak at $q \sim 2 \pi / \sigma$
(which reflects a fluid state of disks with radius $\sigma$). The region of interest of this local structure is represented by a green shaded region in Fig.~\ref{fig:3}a.
}
Specifically, measuring the \blue{relative} height of the peak as the difference between the static structure factor at the maximum of the peak and its value in the nearer minimum found at lower values of $q$ (dotted vertical lines in Fig.~\ref{fig:3}a), one can see that the peak intensity increases by tuning the strength of the Vertex interaction, as shown in panel b) of Fig.~\ref{fig:3}. \blue{This local structure becomes dominant in the disordered solid phase, as shown in the lower row of Fig.~\ref{fig:3}a where, for $s_0=3.3$, the peak at $2\pi/\sigma$ is eventually replaced by the one at $q_V$. In contrast, for $s_0=3.8$ (upper row), the secondary peak remains subdominant, indicating that deep in the fluid regime, although short-range repulsive interactions dominate, the local structure already differs from that of a standard fluid of soft repulsive disks.In other words, due to the progressively dominant effect of geometrical interactions,} cells develop different short-range structures \blue{as qualitatively documented by the representative snapshots in Fig.~\ref{fig:3}c, changing the system from a standard fluid to a Voronoi fluid.}

\section{Nucleus size/hardness drives the Epithelial-to-Mesenchymal transition} 

\blue{The Epithelial-to-Mesenchymal Transition is a fundamental biological process in which epithelial cells undergo biochemical changes that enable them to assume a mesenchymal phenotype. Crucially, cell-cell interactions differ qualitatively between the epithelial and mesenchymal phases.
The mesenchymal phenotype is characterized by enhanced migratory capacity, invasiveness, and greatly reduced adhesion to neighboring cells. From a statistical mechanics point of view, the EMT can be seen as a nonequilibrium phase transition from a confluent, solid-like tissue to a more fluid-like state composed of individual, highly motile cells. In our Voronoi and Vertex models, this transition is captured by tuning 
the shape parameter $s_0$. In the hybrid Voronoi model, the transition is triggered by the competition between vertex interactions, which promote a stable and spatially homogeneous tissue architecture, and nuclear repulsive forces, which dominate when cells act as independent units.}

In Voronoi models the packing fraction (for $N$ cells in a square box of side $L$) is $\varphi=\frac{N \langle A_i\rangle}{L^2}$ and thus $\varphi\!=\!1$. This is because one considers the Voronoi tesselation of $N$ cells in a square box of side $\sqrt{N}$. The ingredient of the cell nucleus introduces the typical length of the nucleus size $\sigma_0$, which allows us to \blue{define the packing fraction $\varphi$ as the fraction of surface covered by the cell nuclei}.
Active particle systems are known to show a rich phase diagram by changing $\varphi$.
The more $\varphi$ approaches its maximum value ($\varphi\!=\!1$), the more cells dispose of in an ordered hexatic phase (see Appendix for details) due to the increased mechanical rigidity of the cells~\cite{Pasupalak2020}. On the other extreme, for low packing fractions, the system assumes the behavior of a gas (see Fig.~\ref{fig:4}a). 
In between, motility-induced phase separation is reported as a function of the persistence time and cell effective radius.   
To assess the effect of cell nuclei on the colony behavior, we explore the phase diagram as a function of $\sigma_0$ and $\epsilon$ while maintaining the other parameters fixed.

Changing $\sigma_0$ produces a rich phase diagram where cells can arrange into gas-like or MIPS at low $\varphi$ (density typical of mesenchymal cells), as one can see from Fig.~\ref{fig:4}b${}_1$ and c1 where we reported the diagrams obtained for $\tau\!=\!50$ and $s_0 = 3.2,3.8$, and $\alpha = 12$.  

By increasing $\epsilon$, the system becomes \blue{structurally uniform, and the density inhomogeneities typical of active fluids are suppressed.} 
However, the structural property of the monolayer changes dramatically with both $\varphi$ and $s_0$ displaying both disordered to ordered solid transitions and \blue{uniform} fluid to disordered solid ones. The latter are phases typically associated with epithelial cells. 
Indeed, by inspecting the typical configurations observed in the different regions of the phase diagram, we can distinguish both the aggregated clusters of cells in the MIPS region (see Fig.~\ref{fig:4}${}_2$-(i) and c${}_2$-(ii)) and the hexagonal disposition of the cells that indicate the presence of a crystal-like behavior of the 2D cell layer ( Fig.~\ref{fig:4}b${}_2$-(iii) and c${}_2$-(i)). The observed behaviors are further confirmed by looking at $S(q)$ (see Fig.~\ref{fig:4}b${}_3$-c${}_3$).
Finally, analyzing the region of the diagram characterized by low values of both $\epsilon$ and $\sigma_0$, we found a transition between a gas-like to a liquid-like phase upon increasing the relative ratio between cell-cell adhesion strength and nuclear size as shown in both the configurations of Fig.~\ref{fig:4}b${}_2$-(ii) and ${}_2$-(iii), and quantified by the shape of the respective structure factors. When cell motility is reduced, transitions occur at lower values of $\epsilon$ (see Figure~\ref{fig:4}d). 
Notably, by varying the nucleus \blue{hardness and the nuclear size}, we observed transitions from crystalline to disordered solid and fluid phases (Figure 4e-f), suggesting \blue{proving that the mechanical property of the nucleus can reshape the collective properties of the tissue.}

\section{Conclusions}
We introduced the hybrid-Voronoi model, able to account for the nuclear \blue{hardness} in influencing the collective behavior of cell colonies. 
Previous research has shown that nuclear rigidity can significantly affect tissue dynamics and morphogenesis, yet integrating this aspect into computational models has often proven challenging. To address this, our model utilizes a hybrid Voronoi approach, enabling the simulation of cells in both mesenchymal and epithelial phases by adjusting the strength of the vertex energy. This approach provides a smooth transition from particle-based to \blue{Voronoi models}, effectively capturing cellular phase behavior without the computational cost and extensive parameter tuning required by phase-field models.

The model requires only a minimal parameterization—namely, the nuclear \blue{hardness}, shape parameter, and motility parameter—allowing for efficient yet realistic simulations of collective cell dynamics. This streamlined parameter set provides a powerful tool for exploring how variations in nuclear \blue{hardness} influence tissue behavior, offering valuable insights into the mechanics underlying tissue organization and transitions.

Using the relative strength between excluded volume forces and Vertex forces, we observe that the latter promote uniform configurations so that MIPS reverses into a homogeneous active fluid and eventually to a disordered solid. As a general result, Vertex interactions tend to destroy any cluster formation. This is a critical feature that might be tested against experiments. 
Moving from fluid to disordered solid \blue{by changing as a relevant control parameter the relative weight between short range repulsion and geometrical interaction}, we observe a liquid-liquid transition from active fluid to a Voronoi fluid. We have documented this behavior through the low-$q$ behavior of $S(q)$ that signals a change in the local structures of the active fluid by varying $\epsilon$.
Changing the packing fraction \blue{$\varphi$ that is the relative surface covered by cell nuclei} we documented several phases: the MIPS phase, a liquid phase, a crystallization region, and a disordered solid state. The small-$\epsilon$ regime of the phase diagram is consistent with the literature of active matter \cite{digregorio2018full,Paoluzzi2022}. 
From a biological viewpoint, our results suggest modulation of nuclear size/stiffness \blue{i.e., the size $\sigma_0$, and the power $\alpha$ of the repulsive potential, respectively} may work as an additional layer of control to tune the mechanical properties of cell colonies besides the known regulation of the apico-basal polarity and cell-cell adhesion. Notably, our findings corroborate the recent observations on the effects of nuclear shape/size in modifying the fluidity of cancer cell organoids~\cite{Grosser2021}.

\blue{In conclusion, our findings suggest that the nucleus acts as a key mechanical regulator for the collective behavior of cell aggregates. We documented that by shifting the balance between nuclear repulsion and cell-cell adhesion, the colony can effectively control a nonequilibrium phase transition, tuning its material properties to meet functional demands like stability or motility.}
As a future direction, it might be interesting to account for alignment interaction that has been shown to play a crucial role in cell rearrangements \cite{Malinverno17,petrolli2019confinement}.

\subsection*{Acknowledgements}
This research was partially funded by grants from ERC-2019-Synergy Grant (ASTRA, n. 855923); EIC-2022-PathfinderOpen (ivBM-4PAP, n. 101098989); Project `National Center for Gene Therapy and Drugs based on RNA Technology' (CN00000041) financed by NextGeneration EU PNRR MUR—M4C2—Action 1.4—Call `Potenziamento strutture di ricerca e creazione di campioni nazionali di R\&S' (CUP J33C22001130001).

\appendix

\section{Simulation details}
We performed Molecular Dynamics simulations of the self-propelled Voronoi model by integrating numerically the stochastic equation for each cell center using the Euler-Maruyama  scheme. 
As an initial condition, we consider a random uniform distribution of $N$ cell centers arranged in a square box of side $L=\sqrt{N}$ with periodic boundary conditions (obtained by replicating the system in every direction).   Next, each simulation has been carried out considering $N=400$ cells. Equations of motion have been integrated with a time step of $dt=10^{-3}$ for $10^7$ steps. 
We monitored that the system reached a stationary state by looking at the behavior of the Vertex energy (see Main Text) and the average shape index $ \langle s \rangle = \frac{1}{N}\sum_i^N \frac{p_i}{\sqrt A_i}$.  Typically, stationary configurations are reached after $\sim 10^4$ steps.   

To compute the forces acting on each cell center due to the Vertex energy term at each time, we evaluate the Voronoi tesselation having cell \blue{nuclear} centers matching the Voronoi region centers. Tesselation was obtained using the Voronoi routine of the Python Scipy library. Given the tesselation, forces can be evaluated analytically, which is the standard fashion for force computation within the Voronoi model of biological tissues.

\section{Forces calculation}
Here, we derived exact analytical expressions for \blue{computing the force given in (\ref{eq:tot_force}).}
\blue{To do so,} we need to explicit the geometrical and repulsion terms, i.e $\nabla_{\uv{r}_i}  E_{v}$ and $ \nabla_{\uv{r}_i}  \Phi$, respectively. 
The geometrical force acting on the nucleus center of cell $i$ is obtained by deriving the total Vertex energy with respect to cell center positions (see for instance ~\cite{Bi2016, Sussman2017, barton2017active}): 
\beq
\vec{F}^i_v = - \frac{dE_v}{d\vec{r}_i} = - \frac{dE_v^i}{d\vec{r}_i} - \sum_{j\in N(i)} \frac{dE_v^j}{d\vec{r}_i}
\eeq
where $N(i)$ represents the nearest neighbours of cell $i$. 
The general term $\frac{dE^{\alpha}_j}{dr^{\alpha}_i}$, i.e. the component $\alpha$ of the variation of the energy of cell i when moving the center of cell j, can be written using the following chain rule:
\beq
\frac{dE^{\alpha}_i}{dr^{\beta}_j} = \sum_\nu \frac{dE^{\alpha}_i}{dh_{ijk}^{\nu}} \frac{dh_{ijk}^{\nu}}{dr_{j}^{\beta}} + \sum_\nu \frac{dE^{\alpha}_i}{dh_{ijl}^{\nu}} \frac{dh_{ijl}^{\nu}}{dr_{j}^{\beta}} 
\eeq
where $\vec{h}_{ijk}$ and $\vec{h}_{ijl}$ are the two shared  \blue{vertices} between cells i and j.  
We have to evaluate the change in energy of a cell upon moving one of its \blue{vertices} and the variation in the position of a vertex when moving the cell center.
The former is given by:
\beq
\frac{dE^{\alpha}_i}{dh_{ijk}^{\nu}} = 2 K_p (p_i -p_0)\frac{dp_i}{dh_{ijk}^{\nu}} + 2 K_A (A_i -A_0)\frac{dA_i}{dh_{ijk}^{\nu}}
\eeq
with 
\beq
\frac{dp_i}{dh_{ijk}^{\nu}} = - (\hat{t}_{kj} + \hat{t}_{ki})
\eeq
and $\vec{t}_{kj} = \vec{h}_{ijk}-  \vec{h}_{kjl}$ and $\vec{t}_{ki} = \vec{h}_{ijk}-  \vec{h}_{lik}$, i.e. the oriented vector having the direction of the edge between the moving vertex ($dh_{ijk}$) and the two  \blue{vertices} that share a cell with it. Note that the directions of the defined vectors are going from the moving vertex toward the other(s).
Similarly, the variation of the area upon moving the vertex is given by:
\beq
\frac{dA_i}{dh_{ijk}^{\nu}} = \frac{1}{2} (l_{ij}\hat{n}_{ij} + l_{ik}\hat{n}_{ik})
\eeq
where $l$ is the length of a given $\vec{t}$ and $\hat{n}$ is the unit vector pointing \blue{outward and} normal to the cell edge.

To obtain an expression for the second term, $\frac{d\vec{h}_{ijk}}{d\vec{r}_{j}}$, we start from the relation that identifies the vertex shared between the cells, i, j, and k:
\beq
\vec{h}_{ijk} = A \vec{r}_i + B \vec{r}_j + \Gamma \vec{r}_k 
\eeq
with the coefficients given by:
\beq
A = || \vec{r}_j - \vec{r}_k ||^2 (\vec{r}_i - \vec{r}_j)\cdot(\vec{r}_i - \vec{r}_k) / D = \frac{|| \vec{r}_{jk} ||^2 }{D} \vec{r}_{ij} \cdot \vec{r}_{ik} 
\eeq

\beq
B = \frac{|| \vec{r}_{ik} ||^2 }{D} \vec{r}_{ji} \cdot \vec{r}_{jk} 
\eeq

\beq
\Gamma = \frac{|| \vec{r}_{ij} ||^2 }{D} \vec{r}_{ki} \cdot \vec{r}_{kj} 
\eeq
and 
\beq
D = 2 || \vec{r}_{ij} \times \vec{r}_{jk}||^2   = 2(r_{ij}^x r_{jk}^y - r_{ij}^y r_{jk}^x)^2
\eeq
(Note that we adopted the definition $\vec{r}_{ij} = \vec{r}_i - \vec{r}_j$)
So we have that:
\beq
\frac{\p \vec{h}_{ijk}}{\p \vec{r}_{j}} = \frac{\p A}{\p \vec{r}_{j}} \vec{r}_{i}  + \frac{\p B}{\p \vec{r}_{j}} \vec{r}_{j}  + \frac{\p \Gamma}{\p \vec{r}_{j}} \vec{r}_{k} +  A  \frac{\p \vec{r}_{i}}{\p \vec{r}_{j}}
\eeq
and 
\begin{multline}
\frac{\p A}{\p r^m_j} = 
\frac{1}{D} \left[ || \vec{r}_j - \vec{r}_k ||^2 (r_i^m - r_k^m + r_i^m-r_j^m) \right] - \frac{1}{D} A \frac{\p D}{\p r_i^m}  
\end{multline}
from which one  can obtain:
\beq
\frac{\p(D A)}{\p r_i^m} = D \frac{\p A}{\p r_i^m} + A \frac{\p D}{\p r_i^m} =  
|| \vec{r}_{jk} ||^2 (r_{ik}^m  + r_{ij}^m)
\eeq
Similarly, we have that
\beq
\frac{\p(D B)}{\p r_i^m} =  2 r_{ik}^m (\vec{r}_{ji} \cdot \vec{r}_{jk}) - ||  \vec{r}_{ik}||^2 r_{jk}^m 
\eeq
 
\beq
\frac{\p(D \Gamma)}{\p r_i^m} =  2 r_{ij}^m (\vec{r}_{ki} \cdot \vec{r}_{kj}) - ||  \vec{r}_{ij}||^2 r_{kj}^m 
\eeq
\beq
\frac{\p D}{\p r_i^m} =  4 \left( r_{ij}^x r_{jk}^y - r_{ij}^y r_{jk}^x \right) \left[ \delta_{mx}r_{jk}^y - \delta_{my} r_{jk}^x \right]
\eeq
Finally,  plugging all together, we get:
\begin{multline}
\frac{\p h^l_{ijk}}{\p r_i^m} = \frac{1}{D}  \left( \frac{\p(DA)}{\p r_i^m} - A \frac{\p D}{\p r_i^m} \right) r_i^l +\\
+ \frac{1}{D}  \left( \frac{\p(DB)}{\p r_i^m} - B \frac{\p D}{\p r_i^m} \right) r_j^l + \\
+ \frac{1}{D}\left( \frac{\p(D\Gamma)}{\p r_i^m} - \Gamma \frac{\p D}{\p r_i^m} \right) r_k^l    + A \delta_{lm} 
\end{multline}
\subsubsection{Repulsive force}
The repulsive force acting on the nucleus center of cell $i$ is obtained by deriving the total repulsive energy  
\beq
\vec{F}^i_r = - \frac{d\Phi_r}{d\vec{r}_i} =  \sum_{j\in N(i)} \frac{d\phi_r^j}{d\vec{r}_i}
\eeq
and 
\beq
\frac{d\Phi_r^j}{dr^\gamma_i} = -\alpha \left( \frac{\sigma_0^{\alpha}}{r_{ij}^{\alpha + 1}}\right) \frac{(r^\gamma_i - r^\gamma_j)}{r_{ij}}
\eeq

\section{Observables}
In the following, we introduce the main descriptors used to characterize the structural and dynamical properties of the system: the static structure factor, the hexatic order parameter, and the diffusion coefficient. The static structure factor quantifies spatial correlations in the cell positions and is defined as
\begin{equation}
S(\mathbf{q}) = \frac{1}{N} \left\langle \sum_{j,k=1}^{N} e^{-i \mathbf{q} \cdot (\mathbf{r}_j - \mathbf{r}_k)} \right\rangle,
\label{eq:S_q}
\end{equation}
where $\mathbf{r}_j$ denotes the position of cell $j$, $\mathbf{q}$ is the wave vector, and $\langle \cdot \rangle$ indicates the average over stationary configurations. 
Note that due to the presence of periodic boundary conditions, physical quantities must be invariant under a translation by the box length $L$ in each spatial direction.
This constraint implies that only a discrete set of wavevectors $\mathbf{q}$ are compatible with the periodicity of the system, namely
\begin{equation}
\mathbf{q} = 2\pi \left( \frac{n_x}{L}, \frac{n_y}{L} \right),
\qquad n_x, n_y \in \mathbb{Z}.
\end{equation}
Therefore, not all $\mathbf{q}$ values are allowed: only these discrete wavevectors can be used when computing $S(\mathbf{q})$ in a periodic simulation box and the smallest accessible spacing in reciprocal space is $\Delta q = \frac{2\pi}{L}$, for the square system of side length $L$. Operatively, $S(q)$ is represented as a function of the magnitude $q = |\mathbf{q}|$, obtained by averaging over wavevectors with the same $|\mathbf{q}|$.
The resulting static structure factor provides information about the degree of spatial order and density fluctuations in the system. In particular, the low-$q$ behavior of $S(q)$ gives information on the behavior of the system at large scale and is used to characterize density fluctuations.
The local hexatic order parameter characterizes the degree of sixfold orientational order around each cell and is defined as 
\begin{equation}
\psi_{6,j} = \frac{1}{N_j} \sum_{k=1}^{N_j} e^{i 6 \theta_{jk}},
\label{eq:psi6_local}
\end{equation}
where the sum runs over the $N_j$ nearest neighbors of particle $j$, and $\theta_{jk}$ is the angle between the bond $\mathbf{r}_k - \mathbf{r}_j$ and a fixed reference axis.  
The global hexatic order parameter is obtained by averaging over all particles:
\begin{equation}
\Psi_6 = \frac{1}{N} \left| \sum_{j=1}^{N} \psi_{6,j} \right|.
\label{eq:psi6_global}
\end{equation}
Large values of $\Psi_6$ indicate strong local orientational order, characteristic of hexatically ordered phases. We used this descriptor to define the region of the phase diagram associated with hexatic crystalline phases.
The diffusion coefficient $D$ quantifies long-time translational motion and is computed from the mean squared displacement (MSD),
\begin{equation}
\mathrm{MSD}(t) =  \frac{1}{N} \sum_{j=1}^{N} \left| \mathbf{r}_j(t) - \mathbf{r}_j^0 \right|^2.
\label{eq:msd}
\end{equation}
  The diffusion coefficient is defined as
\begin{equation}
D = \lim_{t \to \infty} \frac{\mathrm{MSD}(t)}{4t},
\label{eq:D}
\end{equation}
in two spatial dimensions.  
The vanishing of $D_{eff} = \frac{D}{D_0}$ (where $D_0 = v_0^2 /2 \tau$) signals dynamical arrest and the onset of glassy or jammed behavior, as discussed in Ref.~\cite{Bi2016}.
\section{Determination of phase diagrams}
To quantitatively locate the phase boundaries, we performed nonlinear least-squares fits of the order parameters, i.e., the effective diffusion coefficient, the hexatic order parameter, and the static structure factor at $q=0$, as a function of control parameters using a sigmoidal function. 
The control parameters that we used to explore the behavior of the system are the modulus of the cell self-propulsion velocity, $v_0$,  the shape index, $s_0$, nucleus size, $\sigma_0$, and the relative ratio between nuclear and cellular energies, $\epsilon$.
The fitting procedure was implemented in \texttt{Python} using the \texttt{scipy.optimize.curve\_fit} routine from the \texttt{SciPy} library.
The functional form of the fit was chosen as a smooth sigmoid,
\begin{equation}
f(x) = \frac{A_1}{1 + \exp\!\left( -\frac{x - x_c}{\Delta x} \right)},
\label{eq:sigmoid}
\end{equation}
where $A_1$ denotes the asymptotic value of the observable in the two phases, $x_c$ corresponds to the inflection point of the sigmoid (which identifies the phase boundary), and $\Delta x$ characterizes the width of the transition region.
Fitting the simulation data to Eq.~(\ref{eq:sigmoid}) allowed us to extract both the location of the phase boundary $x_c$ and an estimate of the sharpness of the transition. Confidence intervals for the fit parameters were obtained from the covariance matrix returned by the \texttt{curve\_fit} routine. 
Results of the fitting procedures are shown in Figures 5-10.

\begin{figure*}
    \centering
    \includegraphics[width=0.9\linewidth]{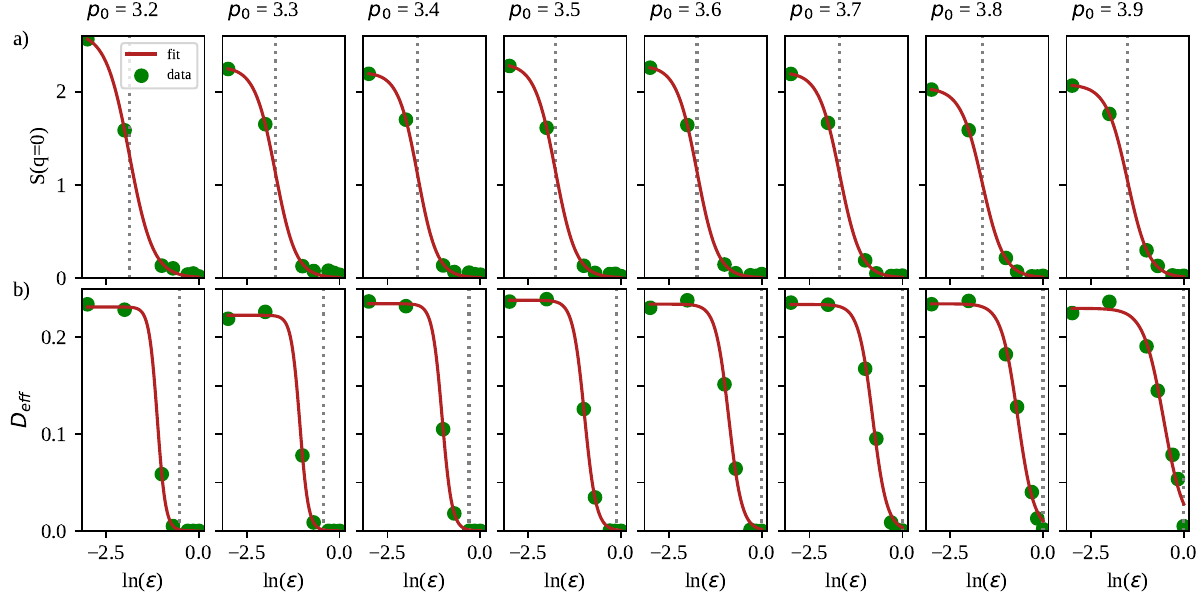}
    \caption{\textbf{Determination of the MIPS and liquid-solid boundary.} \textbf{a)} Static structure factors in $q=0$ as a function of $\epsilon$ for different values of the cells' shape index, $p_0$, used to determine the MIPS boundary shown in Figure~\ref{fig:2}. Green dots represent values obtained from simulations, while the red curve depicts the best-fit solution for a sigmoidal function. Vertical dotted line marks the flex point of the sigmoid.
    \textbf{b)} Effective diffusion coefficient as a function of $\epsilon$ for different values of the cells' shape index, $p_0$, used to determine the solid-liquid boundary shown in Figur~\ref{fig:2}. Green dots represent values obtained from simulations, while the red curve depicts the best-fit solution for a sigmoidal function. Vertical dotted line marks the point where $D_{eff} = 0.001$ in the sigmoid (see Bi \textit{et al.}~\cite{Bi2016}).}
    \label{sfig:s1}
\end{figure*}

\begin{figure*}
    \centering
    \includegraphics[width=0.95\linewidth]{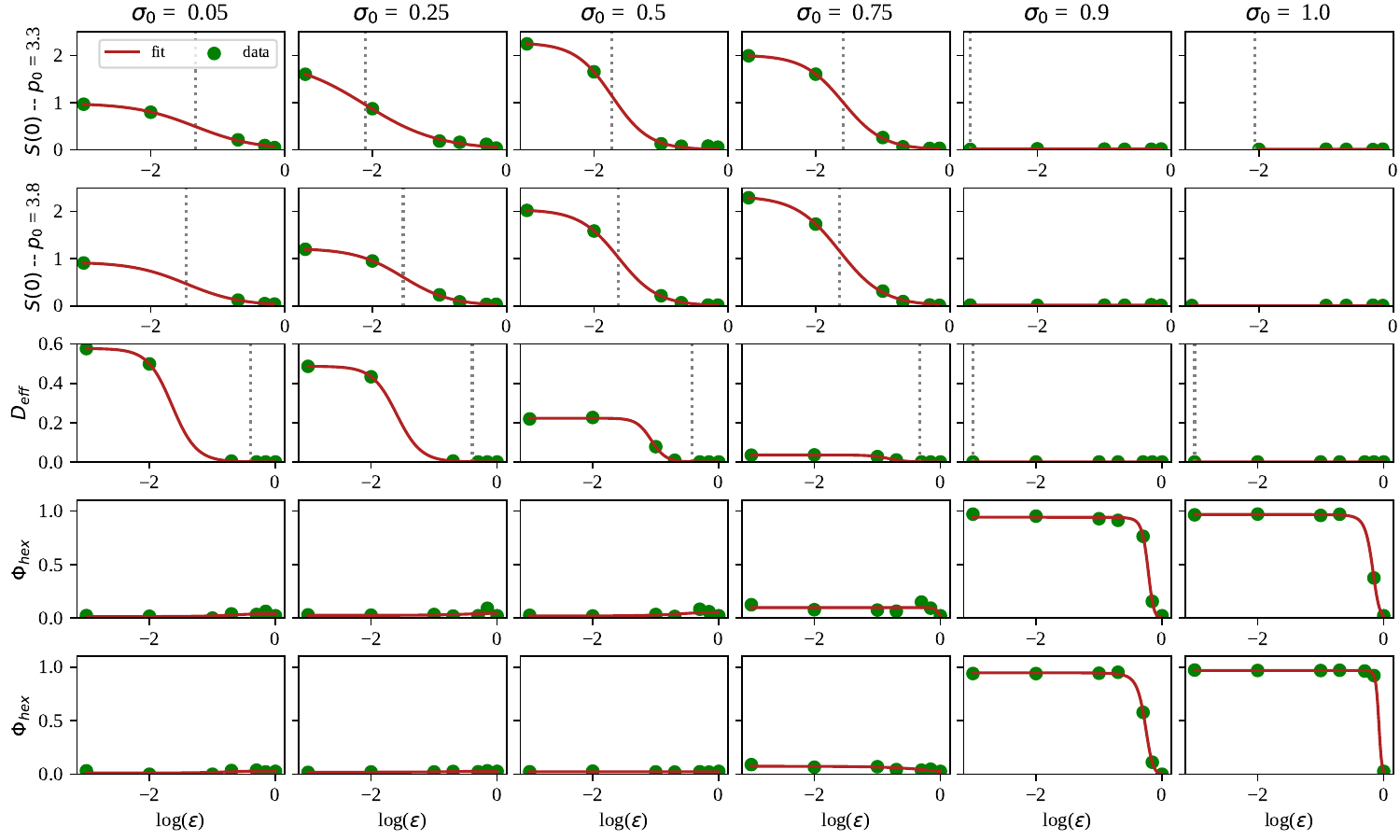}
    \caption{\textbf{Determination of the phase boundaries in $\epsilon$ vs $\sigma_0$ diagrams.} \textbf{a)} Determination of the MIPS boundary. Static structure factors in $q=0$ as a function of $\epsilon$ for different values of the cells' shape index, $p_0$, and nucleus size, $\sigma_0$, used to determine the MIPS boundary shown in Figure 4b. Green dots represent values obtained from simulations, while the red curve depicts the best-fit solution for a sigmoidal function. Vertical dotted line marks the flex point of the sigmoid.
    \textbf{b)} Same as in panel a) but to determine the MIPS boundary shown in Figure 4c.
    \textbf{c)} Determination of the liquid-solid boundary. Effective diffusion coefficient as a function of $\epsilon$ for different values of the cells' nucleus size, $\sigma_0$, used to determine the solid-liquid boundary shown in Figure 4b-1. Green dots represent values obtained from simulations, while the red curve depicts the best-fit solution for a sigmoidal function. Vertical dotted line marks the point where $D_{eff} = 0.001$ in the sigmoid (see Bi \textit{et al.}~\cite{Bi2016}).
    \textbf{d)} Determination of the crystalline boundary. Hexatic order parameter as a function of $\epsilon$ for different values of the cell nucleus size, $\sigma_0$. Trends have been used to determine the crystalline boundaries shown in Figure 4b-1. Green dots represent values obtained from simulations, while the red curve depicts the best-fit solution for a sigmoidal function. Vertical dotted line marks the flex point of the sigmoid.
    \textbf{e)} Same as in panel d) but used to determine the crystalline boundaries shown in Figure 4c-1.
    }
    \label{sfig:s2}
\end{figure*}

\begin{figure*}
    \centering
    \includegraphics[width=0.95\linewidth]{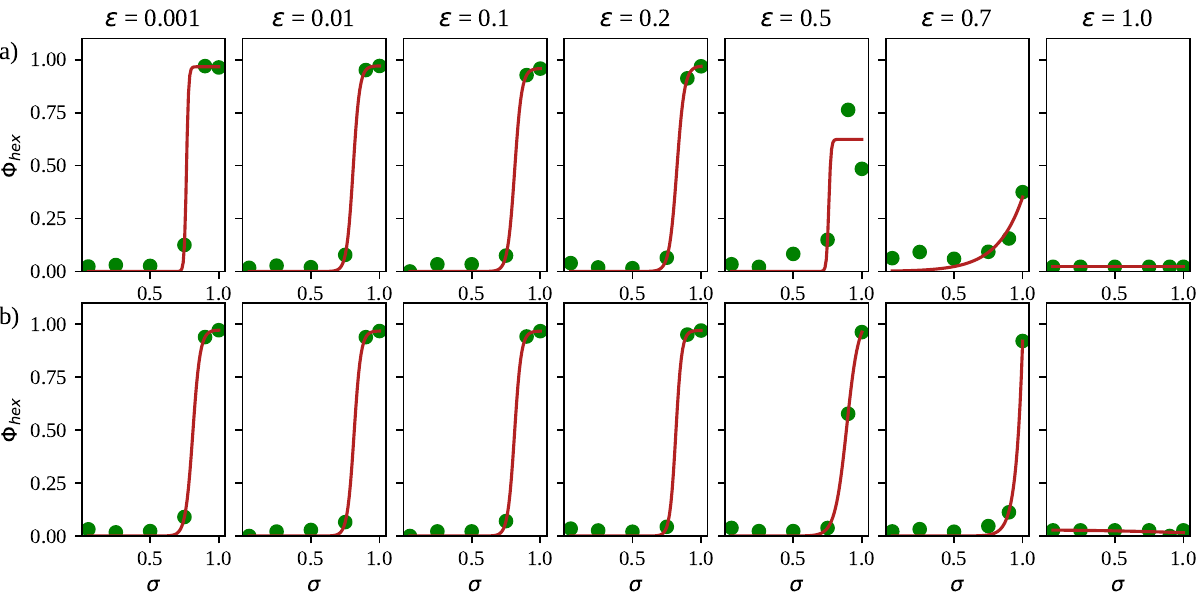}
    \caption{\textbf{Determination of the crystalline boundary in $\epsilon$ vs $\sigma_0$ diagrams.} \textbf{a)} Hexatic order parameter as a function of $\sigma_0$ for different values of $\epsilon$. Trends have been used to determine the crystalline boundaries shown in Figure 4b-1. Green dots represent values obtained from simulations, while the red curve depicts the best-fit solution for a sigmoidal function. Vertical dotted line marks the flex point of the sigmoid.
    \textbf{b)} Same as in panel a) but used to determine the crystalline boundaries shown in Figure 4c-1.}
    \label{sfig:s3}
\end{figure*}

\bibliography{mybibfile.bib}

\providecommand*{\mcitethebibliography}{\thebibliography}
\csname @ifundefined\endcsname{endmcitethebibliography}
{\let\endmcitethebibliography\endthebibliography}{}
\begin{mcitethebibliography}{65}
\providecommand*{\natexlab}[1]{#1}
\providecommand*{\mciteSetBstSublistMode}[1]{}
\providecommand*{\mciteSetBstMaxWidthForm}[2]{}
\providecommand*{\mciteBstWouldAddEndPuncttrue}
  {\def\EndOfBibitem{\unskip.}}
\providecommand*{\mciteBstWouldAddEndPunctfalse}
  {\let\EndOfBibitem\relax}
\providecommand*{\mciteSetBstMidEndSepPunct}[3]{}
\providecommand*{\mciteSetBstSublistLabelBeginEnd}[3]{}
\providecommand*{\EndOfBibitem}{}
\mciteSetBstSublistMode{f}
\mciteSetBstMaxWidthForm{subitem}
{(\emph{\alph{mcitesubitemcount}})}
\mciteSetBstSublistLabelBeginEnd{\mcitemaxwidthsubitemform\space}
{\relax}{\relax}

\bibitem[Alert and Trepat(2020)]{alert2020physical}
R.~Alert and X.~Trepat, \emph{Annual Review of Condensed Matter Physics}, 2020, \textbf{11}, 77--101\relax
\mciteBstWouldAddEndPuncttrue
\mciteSetBstMidEndSepPunct{\mcitedefaultmidpunct}
{\mcitedefaultendpunct}{\mcitedefaultseppunct}\relax
\EndOfBibitem
\bibitem[Trepat and Sahai(2018)]{trepat2018mesoscale}
X.~Trepat and E.~Sahai, \emph{Nature Physics}, 2018, \textbf{14}, 671--682\relax
\mciteBstWouldAddEndPuncttrue
\mciteSetBstMidEndSepPunct{\mcitedefaultmidpunct}
{\mcitedefaultendpunct}{\mcitedefaultseppunct}\relax
\EndOfBibitem
\bibitem[Bi \emph{et~al.}(2016)Bi, Yang, Marchetti, and Manning]{Bi2016}
D.~Bi, X.~Yang, M.~C. Marchetti and M.~L. Manning, \emph{Phys. Rev. X}, 2016, \textbf{6}, 021011\relax
\mciteBstWouldAddEndPuncttrue
\mciteSetBstMidEndSepPunct{\mcitedefaultmidpunct}
{\mcitedefaultendpunct}{\mcitedefaultseppunct}\relax
\EndOfBibitem
\bibitem[Henkes \emph{et~al.}(2020)Henkes, Kostanjevec, Collinson, Sknepnek, and Bertin]{henkes2020dense}
S.~Henkes, K.~Kostanjevec, J.~M. Collinson, R.~Sknepnek and E.~Bertin, \emph{Nature communications}, 2020, \textbf{11}, 1--9\relax
\mciteBstWouldAddEndPuncttrue
\mciteSetBstMidEndSepPunct{\mcitedefaultmidpunct}
{\mcitedefaultendpunct}{\mcitedefaultseppunct}\relax
\EndOfBibitem
\bibitem[Camley and Rappel(2017)]{camley2017physical}
B.~A. Camley and W.-J. Rappel, \emph{Journal of physics D: Applied physics}, 2017, \textbf{50}, 113002\relax
\mciteBstWouldAddEndPuncttrue
\mciteSetBstMidEndSepPunct{\mcitedefaultmidpunct}
{\mcitedefaultendpunct}{\mcitedefaultseppunct}\relax
\EndOfBibitem
\bibitem[Barton \emph{et~al.}(2017)Barton, Henkes, Weijer, and Sknepnek]{barton2017active}
D.~L. Barton, S.~Henkes, C.~J. Weijer and R.~Sknepnek, \emph{PLoS computational biology}, 2017, \textbf{13}, e1005569\relax
\mciteBstWouldAddEndPuncttrue
\mciteSetBstMidEndSepPunct{\mcitedefaultmidpunct}
{\mcitedefaultendpunct}{\mcitedefaultseppunct}\relax
\EndOfBibitem
\bibitem[Li and Ciamarra(2018)]{Li2018}
Y.-W. Li and M.~P. Ciamarra, \emph{Physical Review Materials}, 2018, \textbf{2}, \relax
\mciteBstWouldAddEndPuncttrue
\mciteSetBstMidEndSepPunct{\mcitedefaultmidpunct}
{\mcitedefaultendpunct}{\mcitedefaultseppunct}\relax
\EndOfBibitem
\bibitem[Miotto \emph{et~al.}(2023)Miotto, Rosito, Paoluzzi, de~Turris, Folli, Leonetti, Ruocco, Rosa, and Gosti]{Miotto2023}
M.~Miotto, M.~Rosito, M.~Paoluzzi, V.~de~Turris, V.~Folli, M.~Leonetti, G.~Ruocco, A.~Rosa and G.~Gosti, \emph{Frontiers in Cell and Developmental Biology}, 2023, \textbf{11}, \relax
\mciteBstWouldAddEndPuncttrue
\mciteSetBstMidEndSepPunct{\mcitedefaultmidpunct}
{\mcitedefaultendpunct}{\mcitedefaultseppunct}\relax
\EndOfBibitem
\bibitem[Sunyer \emph{et~al.}(2016)Sunyer, Conte, Escribano, Elosegui-Artola, Labernadie, Valon, Navajas, Garc{\'\i}a-Aznar, Mu{\~n}oz, Roca-Cusachs,\emph{et~al.}]{sunyer2016collective}
R.~Sunyer, V.~Conte, J.~Escribano, A.~Elosegui-Artola, A.~Labernadie, L.~Valon, D.~Navajas, J.~M. Garc{\'\i}a-Aznar, J.~J. Mu{\~n}oz, P.~Roca-Cusachs \emph{et~al.}, \emph{Science}, 2016, \textbf{353}, 1157--1161\relax
\mciteBstWouldAddEndPuncttrue
\mciteSetBstMidEndSepPunct{\mcitedefaultmidpunct}
{\mcitedefaultendpunct}{\mcitedefaultseppunct}\relax
\EndOfBibitem
\bibitem[Friedl and Mayor(2017)]{Friedl17}
P.~Friedl and R.~Mayor, \emph{Cold Spring Harbor perspectives in biology}, 2017, \textbf{9 (4)}, a029199\relax
\mciteBstWouldAddEndPuncttrue
\mciteSetBstMidEndSepPunct{\mcitedefaultmidpunct}
{\mcitedefaultendpunct}{\mcitedefaultseppunct}\relax
\EndOfBibitem
\bibitem[Miotto and Monacelli(2021)]{Miotto2021}
M.~Miotto and L.~Monacelli, \emph{Entropy}, 2021, \textbf{23}, 1138\relax
\mciteBstWouldAddEndPuncttrue
\mciteSetBstMidEndSepPunct{\mcitedefaultmidpunct}
{\mcitedefaultendpunct}{\mcitedefaultseppunct}\relax
\EndOfBibitem
\bibitem[Smeets \emph{et~al.}(2016)Smeets, Alert, Pesek, Pagonabarraga, Ramon, and Vincent]{Smeets16}
B.~Smeets, R.~Alert, J.~Pesek, I.~Pagonabarraga, H.~Ramon and R.~Vincent, \emph{Proceedings of the National Academy of Sciences}, 2016, \textbf{113}, 14621--14626\relax
\mciteBstWouldAddEndPuncttrue
\mciteSetBstMidEndSepPunct{\mcitedefaultmidpunct}
{\mcitedefaultendpunct}{\mcitedefaultseppunct}\relax
\EndOfBibitem
\bibitem[Giavazzi \emph{et~al.}(2017)Giavazzi, Malinverno, Corallino, Ginelli, Scita, and Cerbino]{giavazzi2017giant}
F.~Giavazzi, C.~Malinverno, S.~Corallino, F.~Ginelli, G.~Scita and R.~Cerbino, \emph{Journal of Physics D: Applied Physics}, 2017, \textbf{50}, 384003\relax
\mciteBstWouldAddEndPuncttrue
\mciteSetBstMidEndSepPunct{\mcitedefaultmidpunct}
{\mcitedefaultendpunct}{\mcitedefaultseppunct}\relax
\EndOfBibitem
\bibitem[Paoluzzi \emph{et~al.}(2022)Paoluzzi, Levis, and Pagonabarraga]{Paoluzzi2022}
M.~Paoluzzi, D.~Levis and I.~Pagonabarraga, \emph{Communications Physics}, 2022, \textbf{5}, \relax
\mciteBstWouldAddEndPuncttrue
\mciteSetBstMidEndSepPunct{\mcitedefaultmidpunct}
{\mcitedefaultendpunct}{\mcitedefaultseppunct}\relax
\EndOfBibitem
\bibitem[Paoluzzi \emph{et~al.}(2024)Paoluzzi, Levis, and Pagonabarraga]{paoluzzi2024flocking}
M.~Paoluzzi, D.~Levis and I.~Pagonabarraga, \emph{Communications Physics}, 2024, \textbf{7}, 57\relax
\mciteBstWouldAddEndPuncttrue
\mciteSetBstMidEndSepPunct{\mcitedefaultmidpunct}
{\mcitedefaultendpunct}{\mcitedefaultseppunct}\relax
\EndOfBibitem
\bibitem[Shao \emph{et~al.}(2010)Shao, Rappel, and Levine]{PhysRevLett.105.108104}
D.~Shao, W.-J. Rappel and H.~Levine, \emph{Phys. Rev. Lett.}, 2010, \textbf{105}, 108104\relax
\mciteBstWouldAddEndPuncttrue
\mciteSetBstMidEndSepPunct{\mcitedefaultmidpunct}
{\mcitedefaultendpunct}{\mcitedefaultseppunct}\relax
\EndOfBibitem
\bibitem[Wenzel and Voigt(2021)]{PhysRevE.104.054410}
D.~Wenzel and A.~Voigt, \emph{Phys. Rev. E}, 2021, \textbf{104}, 054410\relax
\mciteBstWouldAddEndPuncttrue
\mciteSetBstMidEndSepPunct{\mcitedefaultmidpunct}
{\mcitedefaultendpunct}{\mcitedefaultseppunct}\relax
\EndOfBibitem
\bibitem[Chiang \emph{et~al.}(2024)Chiang, Hopkins, Loewe, Marchetti, and Marenduzzo]{chiang2024intercellular}
M.~Chiang, A.~Hopkins, B.~Loewe, M.~C. Marchetti and D.~Marenduzzo, \emph{Proceedings of the National Academy of Sciences}, 2024, \textbf{121}, e2319310121\relax
\mciteBstWouldAddEndPuncttrue
\mciteSetBstMidEndSepPunct{\mcitedefaultmidpunct}
{\mcitedefaultendpunct}{\mcitedefaultseppunct}\relax
\EndOfBibitem
\bibitem[Miotto and Monacelli(2018)]{PhysRevE.98.042402}
M.~Miotto and L.~Monacelli, \emph{Phys. Rev. E}, 2018, \textbf{98}, 042402\relax
\mciteBstWouldAddEndPuncttrue
\mciteSetBstMidEndSepPunct{\mcitedefaultmidpunct}
{\mcitedefaultendpunct}{\mcitedefaultseppunct}\relax
\EndOfBibitem
\bibitem[Hopkins \emph{et~al.}(2022)Hopkins, Chiang, Loewe, Marenduzzo, and Marchetti]{PhysRevLett.129.148101}
A.~Hopkins, M.~Chiang, B.~Loewe, D.~Marenduzzo and M.~C. Marchetti, \emph{Phys. Rev. Lett.}, 2022, \textbf{129}, 148101\relax
\mciteBstWouldAddEndPuncttrue
\mciteSetBstMidEndSepPunct{\mcitedefaultmidpunct}
{\mcitedefaultendpunct}{\mcitedefaultseppunct}\relax
\EndOfBibitem
\bibitem[Loewe \emph{et~al.}(2020)Loewe, Chiang, Marenduzzo, and Marchetti]{PhysRevLett.125.038003}
B.~Loewe, M.~Chiang, D.~Marenduzzo and M.~C. Marchetti, \emph{Phys. Rev. Lett.}, 2020, \textbf{125}, 038003\relax
\mciteBstWouldAddEndPuncttrue
\mciteSetBstMidEndSepPunct{\mcitedefaultmidpunct}
{\mcitedefaultendpunct}{\mcitedefaultseppunct}\relax
\EndOfBibitem
\bibitem[Mueller \emph{et~al.}(2019)Mueller, Yeomans, and Doostmohammadi]{PhysRevLett.122.048004}
R.~Mueller, J.~M. Yeomans and A.~Doostmohammadi, \emph{Phys. Rev. Lett.}, 2019, \textbf{122}, 048004\relax
\mciteBstWouldAddEndPuncttrue
\mciteSetBstMidEndSepPunct{\mcitedefaultmidpunct}
{\mcitedefaultendpunct}{\mcitedefaultseppunct}\relax
\EndOfBibitem
\bibitem[Graner and Glazier(1992)]{PhysRevLett.69.2013}
F.~m.~c. Graner and J.~A. Glazier, \emph{Phys. Rev. Lett.}, 1992, \textbf{69}, 2013--2016\relax
\mciteBstWouldAddEndPuncttrue
\mciteSetBstMidEndSepPunct{\mcitedefaultmidpunct}
{\mcitedefaultendpunct}{\mcitedefaultseppunct}\relax
\EndOfBibitem
\bibitem[Sadhukhan and Nandi(2021)]{nandi21}
S.~Sadhukhan and S.~K. Nandi, \emph{Phys. Rev. E}, 2021, \textbf{103}, 062403\relax
\mciteBstWouldAddEndPuncttrue
\mciteSetBstMidEndSepPunct{\mcitedefaultmidpunct}
{\mcitedefaultendpunct}{\mcitedefaultseppunct}\relax
\EndOfBibitem
\bibitem[Belousov \emph{et~al.}(2024)Belousov, Savino, Moghe, Hiiragi, Rondoni, and Erzberger]{PhysRevLett.132.248401}
R.~Belousov, S.~Savino, P.~Moghe, T.~Hiiragi, L.~Rondoni and A.~Erzberger, \emph{Phys. Rev. Lett.}, 2024, \textbf{132}, 248401\relax
\mciteBstWouldAddEndPuncttrue
\mciteSetBstMidEndSepPunct{\mcitedefaultmidpunct}
{\mcitedefaultendpunct}{\mcitedefaultseppunct}\relax
\EndOfBibitem
\bibitem[Glazier and Graner(1993)]{PhysRevE.47.2128}
J.~A. Glazier and F.~m.~c. Graner, \emph{Phys. Rev. E}, 1993, \textbf{47}, 2128--2154\relax
\mciteBstWouldAddEndPuncttrue
\mciteSetBstMidEndSepPunct{\mcitedefaultmidpunct}
{\mcitedefaultendpunct}{\mcitedefaultseppunct}\relax
\EndOfBibitem
\bibitem[Nagai and Honda(2001)]{nagai2001dynamic}
T.~Nagai and H.~Honda, \emph{Philosophical Magazine B}, 2001, \textbf{81}, 699--719\relax
\mciteBstWouldAddEndPuncttrue
\mciteSetBstMidEndSepPunct{\mcitedefaultmidpunct}
{\mcitedefaultendpunct}{\mcitedefaultseppunct}\relax
\EndOfBibitem
\bibitem[Bi \emph{et~al.}(2015)Bi, Lopez, Schwarz, and Manning]{bi2015density}
D.~Bi, J.~Lopez, J.~M. Schwarz and M.~L. Manning, \emph{Nature Physics}, 2015, \textbf{11}, 1074--1079\relax
\mciteBstWouldAddEndPuncttrue
\mciteSetBstMidEndSepPunct{\mcitedefaultmidpunct}
{\mcitedefaultendpunct}{\mcitedefaultseppunct}\relax
\EndOfBibitem
\bibitem[Giavazzi \emph{et~al.}(2018)Giavazzi, Paoluzzi, Macchi, Bi, Scita, Manning, Cerbino, and Marchetti]{giavazzi2018flocking}
F.~Giavazzi, M.~Paoluzzi, M.~Macchi, D.~Bi, G.~Scita, M.~L. Manning, R.~Cerbino and M.~C. Marchetti, \emph{Soft matter}, 2018, \textbf{14}, 3471--3477\relax
\mciteBstWouldAddEndPuncttrue
\mciteSetBstMidEndSepPunct{\mcitedefaultmidpunct}
{\mcitedefaultendpunct}{\mcitedefaultseppunct}\relax
\EndOfBibitem
\bibitem[Merkel and Manning(2018)]{merkel2018geometrically}
M.~Merkel and M.~L. Manning, \emph{New Journal of Physics}, 2018, \textbf{20}, 022002\relax
\mciteBstWouldAddEndPuncttrue
\mciteSetBstMidEndSepPunct{\mcitedefaultmidpunct}
{\mcitedefaultendpunct}{\mcitedefaultseppunct}\relax
\EndOfBibitem
\bibitem[Merkel \emph{et~al.}(2019)Merkel, Baumgarten, Tighe, and Manning]{merkel2019minimal}
M.~Merkel, K.~Baumgarten, B.~P. Tighe and M.~L. Manning, \emph{Proceedings of the National Academy of Sciences}, 2019, \textbf{116}, 6560--6568\relax
\mciteBstWouldAddEndPuncttrue
\mciteSetBstMidEndSepPunct{\mcitedefaultmidpunct}
{\mcitedefaultendpunct}{\mcitedefaultseppunct}\relax
\EndOfBibitem
\bibitem[Erdemci-Tandogan and Manning(2021)]{erdemci2021effect}
G.~Erdemci-Tandogan and M.~L. Manning, \emph{PLoS computational biology}, 2021, \textbf{17}, e1009049\relax
\mciteBstWouldAddEndPuncttrue
\mciteSetBstMidEndSepPunct{\mcitedefaultmidpunct}
{\mcitedefaultendpunct}{\mcitedefaultseppunct}\relax
\EndOfBibitem
\bibitem[Miotto and Monacelli(2020)]{PhysRevResearch.2.043026}
M.~Miotto and L.~Monacelli, \emph{Phys. Rev. Res.}, 2020, \textbf{2}, 043026\relax
\mciteBstWouldAddEndPuncttrue
\mciteSetBstMidEndSepPunct{\mcitedefaultmidpunct}
{\mcitedefaultendpunct}{\mcitedefaultseppunct}\relax
\EndOfBibitem
\bibitem[Yang \emph{et~al.}(2017)Yang, Bi, Czajkowski, Merkel, Manning, and Marchetti]{yang2017correlating}
X.~Yang, D.~Bi, M.~Czajkowski, M.~Merkel, M.~L. Manning and M.~C. Marchetti, \emph{Proceedings of the National Academy of Sciences}, 2017, \textbf{114}, 12663--12668\relax
\mciteBstWouldAddEndPuncttrue
\mciteSetBstMidEndSepPunct{\mcitedefaultmidpunct}
{\mcitedefaultendpunct}{\mcitedefaultseppunct}\relax
\EndOfBibitem
\bibitem[Farhadifar \emph{et~al.}(2007)Farhadifar, R{\"o}per, Aigouy, Eaton, and J{\"u}licher]{farhadifar2007influence}
R.~Farhadifar, J.-C. R{\"o}per, B.~Aigouy, S.~Eaton and F.~J{\"u}licher, \emph{Current biology}, 2007, \textbf{17}, 2095--2104\relax
\mciteBstWouldAddEndPuncttrue
\mciteSetBstMidEndSepPunct{\mcitedefaultmidpunct}
{\mcitedefaultendpunct}{\mcitedefaultseppunct}\relax
\EndOfBibitem
\bibitem[Paoluzzi \emph{et~al.}(2016)Paoluzzi, Di~Leonardo, Marchetti, and Angelani]{paoluzzi2016shape}
M.~Paoluzzi, R.~Di~Leonardo, M.~C. Marchetti and L.~Angelani, \emph{Scientific reports}, 2016, \textbf{6}, 34146\relax
\mciteBstWouldAddEndPuncttrue
\mciteSetBstMidEndSepPunct{\mcitedefaultmidpunct}
{\mcitedefaultendpunct}{\mcitedefaultseppunct}\relax
\EndOfBibitem
\bibitem[Gnan and Zaccarelli(2019)]{gnan2019microscopic}
N.~Gnan and E.~Zaccarelli, \emph{Nature Physics}, 2019, \textbf{15}, 683--688\relax
\mciteBstWouldAddEndPuncttrue
\mciteSetBstMidEndSepPunct{\mcitedefaultmidpunct}
{\mcitedefaultendpunct}{\mcitedefaultseppunct}\relax
\EndOfBibitem
\bibitem[Ton \emph{et~al.}(2024)Ton, MacKeith, Shattuck, and O'Hern]{PhysRevResearch.6.L012036}
A.~T. Ton, A.~K. MacKeith, M.~D. Shattuck and C.~S. O'Hern, \emph{Phys. Rev. Res.}, 2024, \textbf{6}, L012036\relax
\mciteBstWouldAddEndPuncttrue
\mciteSetBstMidEndSepPunct{\mcitedefaultmidpunct}
{\mcitedefaultendpunct}{\mcitedefaultseppunct}\relax
\EndOfBibitem
\bibitem[Park \emph{et~al.}(2015)Park, Kim, Bi, Mitchel, Qazvini, Tantisira, Park, McGill, Kim, Gweon, Jacob~Notbohm, Burger, Randell, Kho, Tambe, Hardin, Shore, Israel, Weitz, Tschumperlin, Henske, Weiss, Manning, Butler, Drazen, and Fredberg]{Park15}
J.-A. Park, J.~H. Kim, D.~Bi, J.~A. Mitchel, N.~T. Qazvini, K.~Tantisira, C.~Y. Park, M.~McGill, S.-H. Kim, B.~Gweon, R.~S.~J. Jacob~Notbohm, S.~Burger, S.~H. Randell, A.~T. Kho, D.~T. Tambe, C.~Hardin, S.~A. Shore, E.~Israel, D.~A. Weitz, D.~J. Tschumperlin, E.~P. Henske, S.~T. Weiss, M.~L. Manning, J.~P. Butler, J.~M. Drazen and J.~J. Fredberg, \emph{Nature Materials}, 2015, \textbf{14}, 1040--1048\relax
\mciteBstWouldAddEndPuncttrue
\mciteSetBstMidEndSepPunct{\mcitedefaultmidpunct}
{\mcitedefaultendpunct}{\mcitedefaultseppunct}\relax
\EndOfBibitem
\bibitem[Arora \emph{et~al.}(2024)Arora, Sadhukhan, Nandi, Bi, Sood, and Ganapathy]{Arora2024}
P.~Arora, S.~Sadhukhan, S.~K. Nandi, D.~Bi, A.~K. Sood and R.~Ganapathy, \emph{Nature Communications}, 2024, \textbf{15}, \relax
\mciteBstWouldAddEndPuncttrue
\mciteSetBstMidEndSepPunct{\mcitedefaultmidpunct}
{\mcitedefaultendpunct}{\mcitedefaultseppunct}\relax
\EndOfBibitem
\bibitem[Sadhukhan \emph{et~al.}(2024)Sadhukhan, Nandi, Pandey, Paoluzzi, Dasgupta, Gov, and Nandi]{Sadhukhan2024}
S.~Sadhukhan, M.~K. Nandi, S.~Pandey, M.~Paoluzzi, C.~Dasgupta, N.~Gov and S.~K. Nandi, \emph{bioRxiv}, 2024\relax
\mciteBstWouldAddEndPuncttrue
\mciteSetBstMidEndSepPunct{\mcitedefaultmidpunct}
{\mcitedefaultendpunct}{\mcitedefaultseppunct}\relax
\EndOfBibitem
\bibitem[Malinverno \emph{et~al.}(2017)Malinverno, Corallino, Giavazzi, Bergert, Li, Leoni, Disanza, Frittoli, Oldani, Martini, Lendenmann, Deflorian, Beznoussenko, Poulikakos, Ong, Uroz, Trepat, Parazzoli, Maiuri, Yu, Ferrari, Cerbino, and Scita]{Malinverno17}
C.~Malinverno, S.~Corallino, F.~Giavazzi, M.~Bergert, Q.~Li, M.~Leoni, A.~Disanza, E.~Frittoli, A.~Oldani, E.~Martini, T.~Lendenmann, G.~Deflorian, G.~V. Beznoussenko, D.~Poulikakos, K.~H. Ong, M.~Uroz, X.~Trepat, D.~Parazzoli, P.~Maiuri, W.~Yu, A.~Ferrari, R.~Cerbino and G.~Scita, \emph{Nature Materials}, 2017, \textbf{16}, 587--596\relax
\mciteBstWouldAddEndPuncttrue
\mciteSetBstMidEndSepPunct{\mcitedefaultmidpunct}
{\mcitedefaultendpunct}{\mcitedefaultseppunct}\relax
\EndOfBibitem
\bibitem[Kang \emph{et~al.}(2021)Kang, Ferruzzi, Spatarelu, Han, Sharma, Koehler, Mitchel, Khan, Butler, Roblyer,\emph{et~al.}]{kang2021novel}
W.~Kang, J.~Ferruzzi, C.-P. Spatarelu, Y.~L. Han, Y.~Sharma, S.~A. Koehler, J.~A. Mitchel, A.~Khan, J.~P. Butler, D.~Roblyer \emph{et~al.}, \emph{Iscience}, 2021, \textbf{24}, \relax
\mciteBstWouldAddEndPuncttrue
\mciteSetBstMidEndSepPunct{\mcitedefaultmidpunct}
{\mcitedefaultendpunct}{\mcitedefaultseppunct}\relax
\EndOfBibitem
\bibitem[Saito and Ishihara(2024)]{Saito2024}
N.~Saito and S.~Ishihara, \emph{Science Advances}, 2024, \textbf{10}, \relax
\mciteBstWouldAddEndPuncttrue
\mciteSetBstMidEndSepPunct{\mcitedefaultmidpunct}
{\mcitedefaultendpunct}{\mcitedefaultseppunct}\relax
\EndOfBibitem
\bibitem[Calero-Cuenca \emph{et~al.}(2018)Calero-Cuenca, Janota, and Gomes]{CaleroCuenca2018}
F.~J. Calero-Cuenca, C.~S. Janota and E.~R. Gomes, \emph{Current Opinion in Cell Biology}, 2018, \textbf{50}, 35–41\relax
\mciteBstWouldAddEndPuncttrue
\mciteSetBstMidEndSepPunct{\mcitedefaultmidpunct}
{\mcitedefaultendpunct}{\mcitedefaultseppunct}\relax
\EndOfBibitem
\bibitem[Boutillon \emph{et~al.}(2024)Boutillon, Banavar, and Campàs]{Boutillon2024}
A.~Boutillon, S.~P. Banavar and O.~Campàs, \emph{Development}, 2024, \textbf{151}, \relax
\mciteBstWouldAddEndPuncttrue
\mciteSetBstMidEndSepPunct{\mcitedefaultmidpunct}
{\mcitedefaultendpunct}{\mcitedefaultseppunct}\relax
\EndOfBibitem
\bibitem[Wolf \emph{et~al.}(2013)Wolf, te~Lindert, Krause, Alexander, te~Riet, Willis, Hoffman, Figdor, Weiss, and Friedl]{Wolf2013}
K.~Wolf, M.~te~Lindert, M.~Krause, S.~Alexander, J.~te~Riet, A.~L. Willis, R.~M. Hoffman, C.~G. Figdor, S.~J. Weiss and P.~Friedl, \emph{Journal of Cell Biology}, 2013, \textbf{201}, 1069–1084\relax
\mciteBstWouldAddEndPuncttrue
\mciteSetBstMidEndSepPunct{\mcitedefaultmidpunct}
{\mcitedefaultendpunct}{\mcitedefaultseppunct}\relax
\EndOfBibitem
\bibitem[Grosser \emph{et~al.}(2021)Grosser, Lippoldt, Oswald, Merkel, Sussman, Renner, Gottheil, Morawetz, Fuhs, Xie, Pawlizak, Fritsch, Wolf, Horn, Briest, Aktas, Manning, and K\"{a}s]{Grosser2021}
S.~Grosser, J.~Lippoldt, L.~Oswald, M.~Merkel, D.~M. Sussman, F.~Renner, P.~Gottheil, E.~W. Morawetz, T.~Fuhs, X.~Xie, S.~Pawlizak, A.~W. Fritsch, B.~Wolf, L.-C. Horn, S.~Briest, B.~Aktas, M.~L. Manning and J.~A. K\"{a}s, \emph{Physical Review X}, 2021, \textbf{11}, \relax
\mciteBstWouldAddEndPuncttrue
\mciteSetBstMidEndSepPunct{\mcitedefaultmidpunct}
{\mcitedefaultendpunct}{\mcitedefaultseppunct}\relax
\EndOfBibitem
\bibitem[Chojowski \emph{et~al.}(2024)Chojowski, Schwarz, and Ziebert]{Chojowski2024}
R.~Chojowski, U.~S. Schwarz and F.~Ziebert, \emph{Soft Matter}, 2024, \textbf{20}, 4488–4503\relax
\mciteBstWouldAddEndPuncttrue
\mciteSetBstMidEndSepPunct{\mcitedefaultmidpunct}
{\mcitedefaultendpunct}{\mcitedefaultseppunct}\relax
\EndOfBibitem
\bibitem[Mitchel \emph{et~al.}(2020)Mitchel, Das, O’Sullivan, Stancil, DeCamp, Koehler, Ocaña, Butler, Fredberg, Nieto, Bi, and Park]{Mitchel2020}
J.~A. Mitchel, A.~Das, M.~J. O’Sullivan, I.~T. Stancil, S.~J. DeCamp, S.~Koehler, O.~H. Ocaña, J.~P. Butler, J.~J. Fredberg, M.~A. Nieto, D.~Bi and J.-A. Park, \emph{Nature Communications}, 2020, \textbf{11}, \relax
\mciteBstWouldAddEndPuncttrue
\mciteSetBstMidEndSepPunct{\mcitedefaultmidpunct}
{\mcitedefaultendpunct}{\mcitedefaultseppunct}\relax
\EndOfBibitem
\bibitem[Tailleur and Cates(2008)]{Tailleur08}
J.~Tailleur and M.~E. Cates, \emph{Phys. Rev. Lett.}, 2008, \textbf{100}, 218103\relax
\mciteBstWouldAddEndPuncttrue
\mciteSetBstMidEndSepPunct{\mcitedefaultmidpunct}
{\mcitedefaultendpunct}{\mcitedefaultseppunct}\relax
\EndOfBibitem
\bibitem[Bechinger \emph{et~al.}(2016)Bechinger, Di~Leonardo, L\"{o}wen, Reichhardt, Volpe, and Volpe]{Bechinger2016}
C.~Bechinger, R.~Di~Leonardo, H.~L\"{o}wen, C.~Reichhardt, G.~Volpe and G.~Volpe, \emph{Reviews of Modern Physics}, 2016, \textbf{88}, \relax
\mciteBstWouldAddEndPuncttrue
\mciteSetBstMidEndSepPunct{\mcitedefaultmidpunct}
{\mcitedefaultendpunct}{\mcitedefaultseppunct}\relax
\EndOfBibitem
\bibitem[Fodor and Marchetti(2018)]{fodor2018statistical}
{\'E}.~Fodor and M.~C. Marchetti, \emph{Physica A: Statistical Mechanics and its Applications}, 2018, \textbf{504}, 106--120\relax
\mciteBstWouldAddEndPuncttrue
\mciteSetBstMidEndSepPunct{\mcitedefaultmidpunct}
{\mcitedefaultendpunct}{\mcitedefaultseppunct}\relax
\EndOfBibitem
\bibitem[Rao and Simha(2013)]{marchetti2013hydrodynamics}
M.~Rao and R.~A. Simha, \emph{Reviews of modern physics}, 2013, \textbf{85}, 1143--1189\relax
\mciteBstWouldAddEndPuncttrue
\mciteSetBstMidEndSepPunct{\mcitedefaultmidpunct}
{\mcitedefaultendpunct}{\mcitedefaultseppunct}\relax
\EndOfBibitem
\bibitem[Hufnagel \emph{et~al.}(2007)Hufnagel, Teleman, Rouault, Cohen, and Shraiman]{hufnagel2007mechanism}
L.~Hufnagel, A.~A. Teleman, H.~Rouault, S.~M. Cohen and B.~I. Shraiman, \emph{Proceedings of the National Academy of Sciences}, 2007, \textbf{104}, 3835--3840\relax
\mciteBstWouldAddEndPuncttrue
\mciteSetBstMidEndSepPunct{\mcitedefaultmidpunct}
{\mcitedefaultendpunct}{\mcitedefaultseppunct}\relax
\EndOfBibitem
\bibitem[Staple \emph{et~al.}(2010)Staple, Farhadifar, R{\"o}per, Aigouy, Eaton, and J{\"u}licher]{staple2010mechanics}
D.~B. Staple, R.~Farhadifar, J.-C. R{\"o}per, B.~Aigouy, S.~Eaton and F.~J{\"u}licher, \emph{The European Physical Journal E}, 2010, \textbf{33}, 117--127\relax
\mciteBstWouldAddEndPuncttrue
\mciteSetBstMidEndSepPunct{\mcitedefaultmidpunct}
{\mcitedefaultendpunct}{\mcitedefaultseppunct}\relax
\EndOfBibitem
\bibitem[Hilgenfeldt \emph{et~al.}(2008)Hilgenfeldt, Erisken, and Carthew]{hilgenfeldt2008physical}
S.~Hilgenfeldt, S.~Erisken and R.~W. Carthew, \emph{Proceedings of the National Academy of Sciences}, 2008, \textbf{105}, 907--911\relax
\mciteBstWouldAddEndPuncttrue
\mciteSetBstMidEndSepPunct{\mcitedefaultmidpunct}
{\mcitedefaultendpunct}{\mcitedefaultseppunct}\relax
\EndOfBibitem
\bibitem[Manning \emph{et~al.}(2010)Manning, Foty, Steinberg, and Schoetz]{manning2010coaction}
M.~L. Manning, R.~A. Foty, M.~S. Steinberg and E.-M. Schoetz, \emph{Proceedings of the National Academy of Sciences}, 2010, \textbf{107}, 12517--12522\relax
\mciteBstWouldAddEndPuncttrue
\mciteSetBstMidEndSepPunct{\mcitedefaultmidpunct}
{\mcitedefaultendpunct}{\mcitedefaultseppunct}\relax
\EndOfBibitem
\bibitem[Wang \emph{et~al.}(2012)Wang, Manning, and Amack]{wang2012regional}
G.~Wang, M.~L. Manning and J.~D. Amack, \emph{Developmental biology}, 2012, \textbf{370}, 52--62\relax
\mciteBstWouldAddEndPuncttrue
\mciteSetBstMidEndSepPunct{\mcitedefaultmidpunct}
{\mcitedefaultendpunct}{\mcitedefaultseppunct}\relax
\EndOfBibitem
\bibitem[Chiou \emph{et~al.}(2012)Chiou, Hufnagel, and Shraiman]{chiou2012mechanical}
K.~K. Chiou, L.~Hufnagel and B.~I. Shraiman, \emph{PLoS computational biology}, 2012, \textbf{8}, e1002512\relax
\mciteBstWouldAddEndPuncttrue
\mciteSetBstMidEndSepPunct{\mcitedefaultmidpunct}
{\mcitedefaultendpunct}{\mcitedefaultseppunct}\relax
\EndOfBibitem
\bibitem[Fletcher \emph{et~al.}(2014)Fletcher, Osterfield, Baker, and Shvartsman]{fletcher2014vertex}
A.~G. Fletcher, M.~Osterfield, R.~E. Baker and S.~Y. Shvartsman, \emph{Biophysical journal}, 2014, \textbf{106}, 2291--2304\relax
\mciteBstWouldAddEndPuncttrue
\mciteSetBstMidEndSepPunct{\mcitedefaultmidpunct}
{\mcitedefaultendpunct}{\mcitedefaultseppunct}\relax
\EndOfBibitem
\bibitem[Pasupalak \emph{et~al.}(2020)Pasupalak, Yan-Wei, Ni, and Pica~Ciamarra]{Pasupalak2020}
A.~Pasupalak, L.~Yan-Wei, R.~Ni and M.~Pica~Ciamarra, \emph{Soft Matter}, 2020, \textbf{16}, 3914–3920\relax
\mciteBstWouldAddEndPuncttrue
\mciteSetBstMidEndSepPunct{\mcitedefaultmidpunct}
{\mcitedefaultendpunct}{\mcitedefaultseppunct}\relax
\EndOfBibitem
\bibitem[Digregorio \emph{et~al.}(2018)Digregorio, Levis, Suma, Cugliandolo, Gonnella, and Pagonabarraga]{digregorio2018full}
P.~Digregorio, D.~Levis, A.~Suma, L.~F. Cugliandolo, G.~Gonnella and I.~Pagonabarraga, \emph{Physical review letters}, 2018, \textbf{121}, 098003\relax
\mciteBstWouldAddEndPuncttrue
\mciteSetBstMidEndSepPunct{\mcitedefaultmidpunct}
{\mcitedefaultendpunct}{\mcitedefaultseppunct}\relax
\EndOfBibitem
\bibitem[Petrolli \emph{et~al.}(2019)Petrolli, Le~Goff, Tadrous, Martens, Allier, Mandula, Herv{\'e}, Henkes, Sknepnek, Boudou,\emph{et~al.}]{petrolli2019confinement}
V.~Petrolli, M.~Le~Goff, M.~Tadrous, K.~Martens, C.~Allier, O.~Mandula, L.~Herv{\'e}, S.~Henkes, R.~Sknepnek, T.~Boudou \emph{et~al.}, \emph{Physical review letters}, 2019, \textbf{122}, 168101\relax
\mciteBstWouldAddEndPuncttrue
\mciteSetBstMidEndSepPunct{\mcitedefaultmidpunct}
{\mcitedefaultendpunct}{\mcitedefaultseppunct}\relax
\EndOfBibitem
\bibitem[Sussman(2017)]{Sussman2017}
D.~M. Sussman, \emph{Computer Physics Communications}, 2017, \textbf{219}, 400–406\relax
\mciteBstWouldAddEndPuncttrue
\mciteSetBstMidEndSepPunct{\mcitedefaultmidpunct}
{\mcitedefaultendpunct}{\mcitedefaultseppunct}\relax
\EndOfBibitem
\end{mcitethebibliography}
\bibliographystyle{rsc}

\end{document}